\documentclass[9pt,sigconf,letterpaper]{acmart}
\usepackage{color}
\usepackage{graphicx}
\usepackage[labelformat=simple]{subcaption}
\usepackage{xspace}
\usepackage{multirow}
\usepackage[ruled,vlined]{algorithm2e}
\usepackage{ulem}
\usepackage{xurl}
\usepackage{enumitem}

\usepackage{seqsplit}
\usepackage{amsthm}
\usepackage{geometry}
\usepackage{thmtools} 
\usepackage{tabularx}
\usepackage{booktabs}
\usepackage{balance}

\usepackage[table]{xcolor}

\newcommand{\colmin}[1]{\cellcolor{gray!45}{#1}}   

\declaretheoremstyle[
    style=definition, 
    headfont=\normalfont,
    spaceabove=6pt,
    spacebelow=6pt, 
    bodyfont=\normalfont
]{compact}

\setlist[itemize]{topsep=6pt, itemsep=0pt, parsep=3pt}
\setlist[enumerate]{topsep=6pt, itemsep=0pt, parsep=3pt}

\copyrightyear{2026}
\acmYear{2026}
\setcopyright{cc}
\setcctype{by}
\acmConference[IMC '26]{Proceedings of the 2026 ACM Internet Measurement Conference}{October 12--16, 2026}{Karlsruhe, Germany}
\acmBooktitle{Proceedings of the 2026 ACM Internet Measurement Conference (IMC '26), October 12--16, 2026, Karlsruhe, Germany}
\acmDOI{10.1145/3777912.3809156}
\acmISBN{979-8-4007-2327-8/2026/10}

\begin{document}

\title{Behavioral Consistency and Transparency Analysis on Large Language Model API Gateways}

\author{Guanjie Lin}
\affiliation{%
  \institution{University of Massachusetts Boston}
  \city{Boston}
  \state{Massachusetts}
  \country{USA}}
\email{guanjie.lin001@umb.edu}

\author{Yinxin Wan}
\affiliation{%
  \institution{University of Massachusetts Boston}
  \city{Boston}
  \state{Massachusetts}
  \country{USA}}
\email{yinxin.wan@umb.edu}

\author{Shichao Pei}
\affiliation{%
  \institution{University of Massachusetts Boston}
  \city{Boston}
  \state{Massachusetts}
  \country{USA}}
\email{shichao.pei@umb.edu}

\author{Ting Xu}
\affiliation{%
  \institution{University of Massachusetts Boston}
  \city{Boston}
  \state{Massachusetts}
  \country{USA}}
\email{ting.xu001@umb.edu}

\author{Kuai Xu}
\affiliation{%
  \institution{Arizona State University}
  \city{Glendale}
  \state{Arizona}
  \country{USA}}
\email{kuai.xu@asu.edu}

\author{Guoliang Xue}
\affiliation{%
  \institution{Arizona State University}
  \city{Tempe}
  \state{Arizona}
  \country{USA}}
\email{xue@asu.edu}


\keywords{LLM API gateways, black-box auditing, trustworthy AI, LLM model fingerprinting, API reliability}


\begin{abstract}
    Third-party Large Language Model (LLM) API gateways are rapidly emerging as unified access points to models offered by multiple vendors.
    However, the internal routing, caching, and billing policies of these gateways are largely undisclosed, leaving users with limited visibility into whether requests are served by the advertised models, whether responses remain faithful to upstream APIs, or whether invoices accurately reflect public pricing policies.
    To address this gap, we introduce 
    GateScope, a lightweight black-box measurement framework for evaluating behavioral consistency and operational transparency in commercial LLM gateways.
    GateScope is designed to detect key misbehaviors, including model downgrading or switching, silent truncation, billing inaccuracies, and instability in latency by auditing gateways along four critical dimensions: response content analysis, multi-turn conversation performance, billing accuracy, and latency characteristics.
    Our measurements across $10$ real-world commercial LLM API gateways reveal frequent gaps between expected and actual behaviors, including silent model substitutions, degraded memory retention, deviations from announced pricing, and substantial variation in latency stability across platforms.
\end{abstract}

\ccsdesc[500]{Networks~Network measurement}
\ccsdesc[300]{Security and privacy~Software security engineering}
\ccsdesc[500]{Networks~Network performance analysis}

\maketitle


\section{Introduction}\label{sec:intro}
\noindent
Large Language Models (LLMs) such as OpenAI’s GPT, Anthropic’s Claude, and Google’s Gemini are increasingly integrated into a broad range of applications~\cite{AIIndex2025,Hao:IMC2025}. 
Although these models are widely accessible through cloud-based inference APIs, each vendor maintains its own independent API platform. 
For developers and enterprise users who require access to multiple models from different providers, this fragmented ecosystem introduces substantial operational overhead, including complex backend integration, API key management, and billing processes~\cite{Jimmy:web2025}.

LLM gateways are emerging as a promising solution, offering a unified API interface with centralized billing that aggregates access to models from multiple vendors.
In such gateways, a single API key can be used to invoke different models, significantly simplifying deployment workflows and reducing integration and key management complexity.
In addition, LLM API gateways often offer additional benefits such as load balancing across models, usage-based optimization, and in some cases discounted token pricing.
However, from the client’s perspective, these gateways operate as \emph{black-box} intermediaries with little visibility into their internal behaviors. 
Thus, users cannot reliably verify whether requests use the intended model, whether conversational context is preserved across interactions, or whether billing is accurate~\cite{Cheng:IMC25,Reddit:web2025,Sun:preprint25,Wang:preprint25,Shanmugarasa:SoKPrivacy25}.

Auditing these black-box gateways is challenging because their routing and model selection behaviors are not exposed to clients.
Existing efforts such as LLMmap~\cite{Pasquini:SEC25} use active probing to differentiate between LLMs for mostly open-source models, but do not target the behavioral transparency of commercial LLM gateways.
Other work on instructional fingerprints~\cite{Xu:NAACL24,Yang:FingerLLM24,Song:CVPR24}, UTF~\cite{Cai:ACL25}, AuditLLM~\cite{Amirizaniani:AuditLLM24}, and FDLLM~\cite{Fu:FDLLM25} focuses on analyzing model traits or downstream artifacts rather than evaluating whether third-party API gateways preserve upstream model behavior or comply with pricing rules.
These gaps motivate an efficient framework that measures the transparency and behavioral consistency of LLM gateways from the client’s perspective.

Building on these needs, we present GateScope, a systematic framework for auditing gateway behavior transparency and consistency across four key dimensions.
GateScope performs response content analysis using a structured set of probing queries to generate behavioral profiles for different models, enabling the detection of potential model downgrading or substitution.
It evaluates multi-turn conversational consistency to identify model switching during extended conversations and to determine whether silent truncation occurs before reaching documented context or output limits.
GateScope further examines billing accuracy by comparing actual token consumption with reported usage and assessing differences between expected and reported costs, accounting for cached tokens.
Finally, it measures latency characteristics, quantifying variability in request and response delays.

We first demonstrate GateScope’s effectiveness and efficiency through controlled validation using official vendor model endpoints.
We then apply GateScope to $10$ commercial LLM API gateways, where we uncover significant gaps between expected and observed behaviors, including model substitutions, degraded memory retention, inaccurate billing, and substantial variability in request latency. Our artifacts have been released on GitHub~\cite{GateScope:GitHub}.

Our key research contributions are summarized as follows:

\begin{itemize}[leftmargin=*, topsep=3pt, itemsep=0pt, parsep=3pt]
    \item We identify the lack of transparency and consistency guarantees in LLM API gateways and motivate the need for systematic client-side auditing.
    \item We design GateScope, a lightweight framework that measures gateway behavior across four dimensions: response content, multi-turn conversation consistency, billing accuracy, and latency characteristics.
    \item We validate GateScope in controlled settings and apply it to $10$ commercial gateways, uncovering substantial discrepancies between expected and observed behavior. 
\end{itemize}


\begin{figure}[h]
    \centering
    \vspace{-0.12in}
    \includegraphics[width=2.35in]{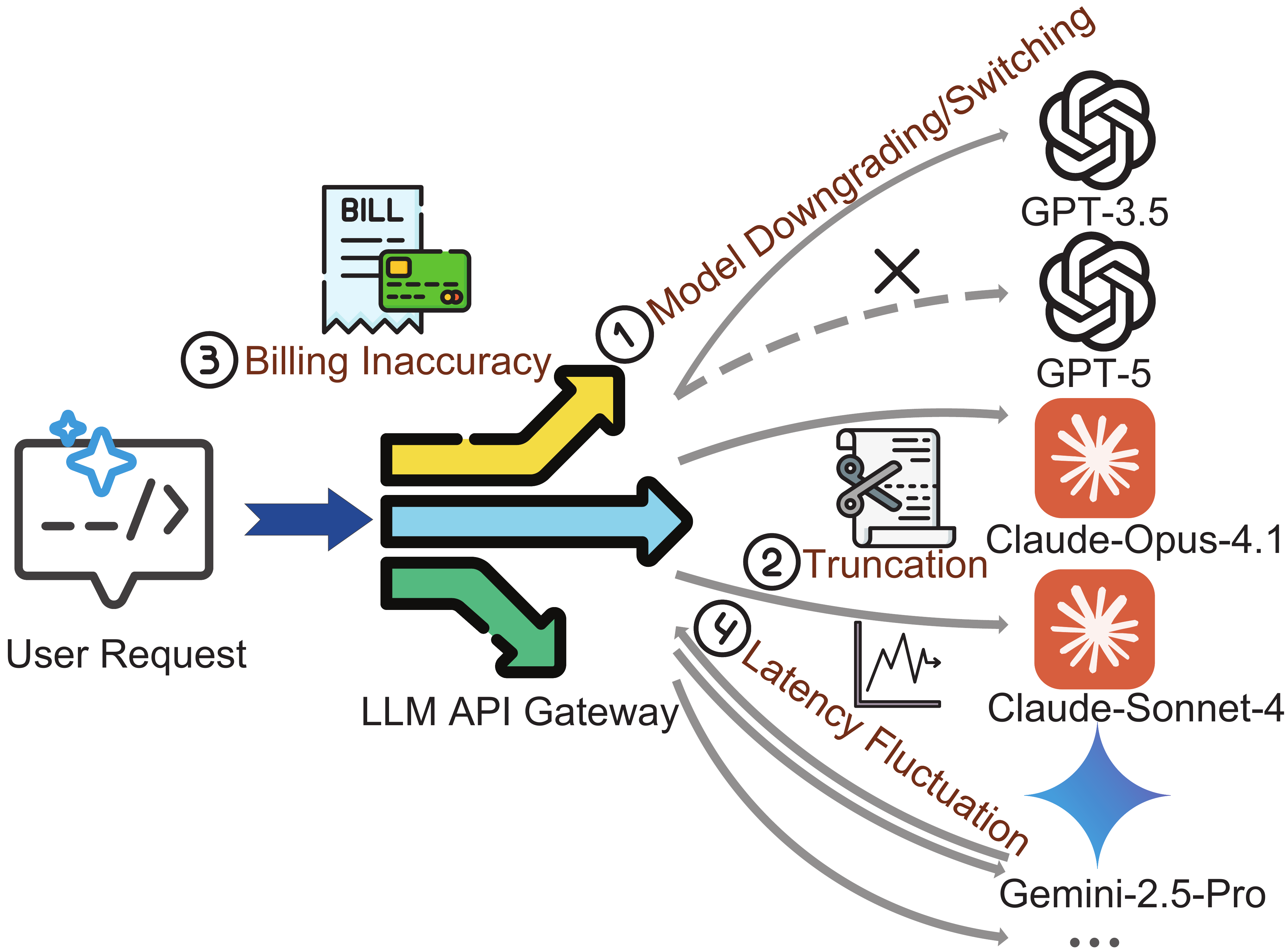}
    \vspace{-0.08in}
    \caption{Overview of LLM API gateway architecture and the main transparency and consistency challenges.}\label{fig:gateway-cheat}\vspace{-0.25in}
\end{figure}

\section{Background and Motivation}\label{sec:background}
LLM API gateways are increasingly adopted to abstract vendor differences and provide unified interfaces, load balancing, and centralized billing~\cite{Openrouter:web2025}.
As illustrated in Figure~\ref{fig:gateway-cheat}, a gateway acts as a proxy between client applications and upstream model providers.
However, the internal operation of these gateways remains largely non-transparent.
Users cannot observe how requests are routed, which upstream models are selected, what inference parameters are applied, or how billing is computed.
As a result, clients cannot determine whether a gateway follows documented policies or behaves consistently. 
This lack of visibility raises concerns around several critical issues as shown in Figure~\ref{fig:gateway-cheat}:

\begin{itemize}[leftmargin=*, topsep=3pt, itemsep=0pt, parsep=3pt]
    \item Model downgrading/switching: 
    gateways may route requests to cheaper or less capable models despite advertising a premium model, often driven by internal cost or latency considerations.
    Such routing decisions are typically undisclosed, leaving users unaware of which model actually served a request.
    
    \item
    Prompt/response truncation: gateways may impose their own context window or maximum output token limits that are smaller than the underlying model's native capacity. When a request exceeds these gateway-imposed limits, the gateway may silently discard portions of the multi-turn prompt history or truncate the model's response before forwarding it to the user. We use the term \emph{silent truncation} throughout this paper to refer specifically to the case where a gateway drops part of the prompt history or response due to its own hidden limits, without explicitly signaling this behavior to the users, thereby silently degrading service quality. This is distinct from the underlying model itself reaching its native boundary, which is typically documented and observable through the API's usage metadata.

    \item Billing inaccuracy: 
    reported usage and actual token consumption may diverge, including cases where gateways charge for cached tokens~\cite{OpenAI:web,Gemini:web} or add extra tokens that were not processed.

    \item Latency fluctuation: limited computational resources, especially during peak hours, along with extra processing and forwarding steps inside the gateway, as well as behaviors such as switching models, modifying user input, or changing API endpoints, can introduce significant variability in response latency at LLM gateways, negatively impacting service quality and reliability.
\end{itemize}

Prior work has examined hidden behaviors in LLM services, such as prompt caching, hidden token usage, and API-level security risks, but these studies primarily focus on first-party vendors' LLM API platforms rather than third-party LLM API gateways~\cite{Gu:PromptCache25,Hyun:ICST24,Sun:preprint25,Wang:preprint25,Wu:SEC25,Shanmugarasa:SoKPrivacy25,Yao:OSDI24}. 
These challenges collectively motivate GateScope, a framework for systematically auditing behavioral transparency and consistency in black-box LLM API gateways.


\section{Design of GateScope}\label{sec:design}
This section describes the design of GateScope in detail. 
The implementation artifacts are publicly available on GitHub~\cite{GateScope:GitHub}.

\begin{figure*}[htbp]
\centering
\includegraphics[width=5.5in]{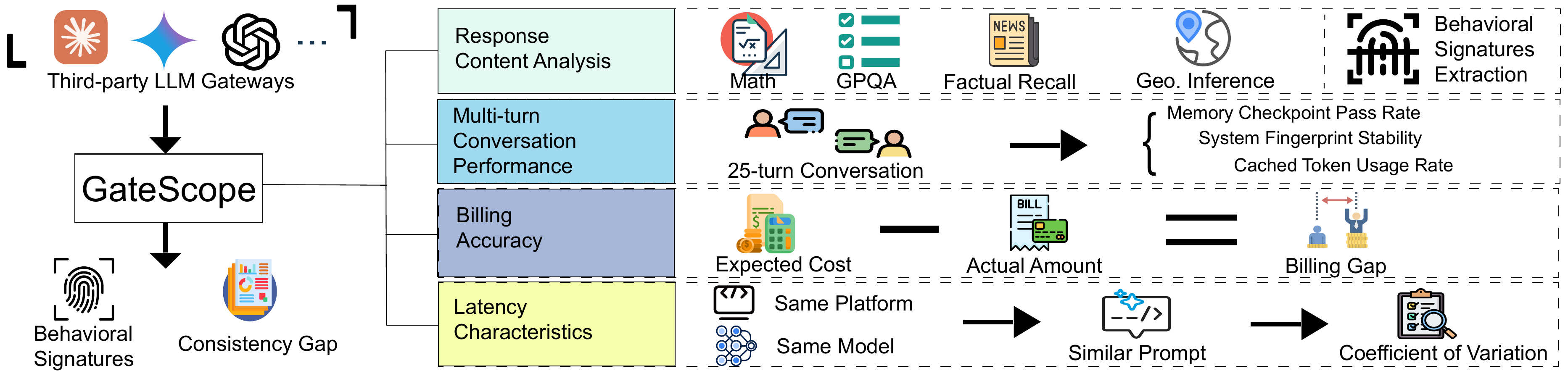}
\caption{Overall architecture of the GateScope framework.}\label{fig:architecture}\vspace{-0.2in}
\end{figure*}

\subsection{Overview of GateScope}\label{sec:design-overview}
\noindent
GateScope is a lightweight, systematic auditing framework that operates entirely through the public APIs exposed by LLM gateways.
All measurements are performed using the OpenAI-compatible \texttt{/v1/chat/completions} endpoint, ensuring a consistent request– response format across platforms.
The framework relies solely on information accessible to ordinary clients through standard interactions, without requiring privileged access or internal instrumentation.
Accordingly, GateScope focuses on detecting observable consistency gaps and supporting signals across the public API surface, rather than inferring internal implementation details.

Figure~\ref{fig:architecture} illustrates the overall structure of GateScope, which implements a multi-dimensional auditing pipeline to evaluate the response content, multi-turn conversation, billing accuracy, and latency characteristics.

\subsection{Response Content Analysis}\label{sec:content-analysis}
\noindent
As shown in Figure~\ref{fig:architecture}, this component focuses on analyzing the behavior of LLM API gateways by examining the response content produced under different sets of input prompts.
The goal of this dimension is to determine whether a gateway actually serves the requested model or silently replaces it with a different and often cheaper model.
To detect such substitution, we analyze the response behavior of different LLMs and use it to differentiate between models.
This is challenging because even the same model may generate varied responses to the same prompt, especially for open-ended questions, making naive text matching unreliable.
Prior work on LLM fingerprinting and identification primarily distinguishes models using lexical, semantic, or logit-space features extracted from free-form text~\cite{Pasquini:SEC25,Zhou:ACL24}.
However, these approaches are often sensitive to generation randomness and frequently fail to capture meaningful behavioral differences in critical fields (e.g., the final answer in a math query is far more important than the surrounding derivation text) when relying solely on text-level analysis.

Instead, our prompt design follows three requirements: (1) responses to the same probe should remain stable across repeated queries to the same model,
(2) responses should differ across models because they reflect model-specific capabilities, and these differences should be identifiable through key feature dimensions, and
(3) the prompt suite should be lightweight so that auditing remains efficient.

To formalize this analysis, we adopt a formulation following prior work~\cite{Pasquini:SEC25,Zhou:ACL24}, treating each model version $v$ as a stochastic mapping
$u \sim E_v(p,\theta)$,
where a probe $p\in\mathcal{P}$ and generation parameters $\theta$ produce an output $u$ with metadata~$m$.

\subsubsection{Prompt Design.}
Following the three requirements mentioned above, we design prompts that (i) structure model responses for analysis and (ii) expose key response features that can effectively differentiate models.
Each prompt instructs the LLM to generate explicit step-by-step reasoning leading to a final answer, and the response is required to follow a shared JSON schema that enforces a consistent output format across models and runs.
We use the shared JSON schema for two reasons. First, it reduces dependence on free-form surface text by forcing responses into a consistent, observable structure, minimizing variation introduced by post-processing differences. Second, it makes the reasoning path and final answer explicit, enabling stable extraction of behavioral features such as reasoning depth, step length, parsing success, and answer placement across models and repeated runs.

When designing the prompt content, we deliberately avoid two unsuitable choices.
First, we avoid banner-grabbing questions such as “What is your model version?”, since they are easily spoofed through system-prompt manipulation and yield inconsistent identifiers across gateways~\cite{Pasquini:SEC25}.
Second, we avoid using large public benchmarks at API scale, as they are expensive to run and risk data contamination~\cite{Hendrycks:MMLU2020,Dan:NIPS2021}.

Motivated by these requirements and constraints, we design a probe suite $\mathcal{S}$ that captures distinctive behavioral signatures across models, including differences in reasoning ability, knowledge cutoff, mathematical competence, and output style (e.g., preferred symbols or \LaTeX~formatting).
To cover these dimensions, we design prompts across four domains:
 (1) \emph{AIME (American Invitational Mathematics Examination)}: numeric-answer problems that test precise computation and mathematical reasoning~\cite{Liu:ACL2025};
 (2) \emph{GPQA (Graduate-Level Google-Proof Q\&A)}: multiple-choice questions designed to expose deep reasoning and deliberation capabilities across models~\cite{Hugging:GPQA2025};
 (3) \emph{Factual Recall}: questions about major events between 2021–2025 that assess knowledge cutoff and factual retrieval ability;
 (4) \emph{Geographic Information Inference}: implicit reasoning tasks that require step-by-step linking of intermediate facts to identify a specific geographic feature.
These four domains provide complementary signals for behavioral-signature extraction: AIME emphasizes precise mathematical reasoning, GPQA targets deliberation depth, factual recall exposes knowledge-cutoff boundaries, and geographic inference captures multi-hop composition style. To ensure that responses reflect only the model's inherent knowledge rather than externally retrieved content, we disable network search and all retrieval-augmented options in the API requests.

Building on discrete concept composition (DCC)~\cite{Guan:Latent2025},
we design each prompt to instruct the LLM to compose disconnected facts
across multiple reasoning hops, making intermediate choices such as
ordering, granularity, and error types directly observable.
Concretely, each response is required to follow a structured format with two fields:
knowledge\_path (an array of intermediate reasoning steps) and
\seqsplit{final\_answer} (the final result).
A complete specification of the request template can be found in  Appendix~\ref{app:dcc-template}.
This structure helps to remove most free-form variations in responses so that observable differences mainly reflect model capabilities rather than surface text patterns.
It also exposes reasoning strategies that remain stable across repetitions.

\subsubsection{Model Signature Extraction.}
For each structured response obtained from a prompt, we extract a $d$-dimensional signature vector $\Phi(u,m)$ rather than relying on text-level embeddings.
This enables robust differentiation between models by capturing behavioral and structural characteristics.
The signature vector is composed of five feature families:
(i) answer quality: correctness of the final answer and whether it appears in the correct location,
(ii) reasoning structure: reasoning depth, mean step length, and variance, 
(iii) scale: response length and density, 
(iv) style: presence of numeric expressions or \LaTeX\ formatting, and 
(v) parsing quality: whether parsing occurs and at what degree. 
Full feature definitions can be found in Appendix~\ref{app:feature}.
For each model, collecting signatures across multiple probes, each repeated several times, yields a behavioral signature matrix whose rows correspond to individual probe repetitions and whose columns correspond to extracted features.
Using these signatures, we train lightweight per-model classifiers. For each model $v$, a one-vs-rest XGBoost classifier is trained to distinguish signatures produced by $v$ from those of all other models.

\subsection{Multi-turn Conversation Performance}\label{sec:conversation}
\noindent
As illustrated in Figure~\ref{fig:architecture}, this component focuses on analyzing the behavior of LLM API gateways when handling long, multi-turn conversations with the user.
LLM applications typically involve multi-turn conversations, where users expect the model to retain context, chat history, and intermediate results across turns.
Since most LLMs are inherently stateless, the full conversation history is included in every request to preserve context in multi-turn interactions.
Under normal conditions, this enables consistent behavior and correct recall of information introduced earlier in the dialog.
However, a black-box gateway can silently break these expectations by switching back-end models to cheaper alternatives or performing silent truncation, either discarding parts of the conversation history or cutting down the model's output, to lower token consumption.

In this paper, we use multi-turn conversational consistency to denote whether a gateway preserves the underlying model's claimed input and output token limits, maintaining full conversation context and complete responses as the dialog continues across turns. Within this dimension, we focus on context loss caused by hidden context-management limits, where a gateway silently drops part of the prompt history or truncates the response without signaling this behavior to the client.
We evaluate gateway behavior using a multi-turn procedure with repeated questions and explicit checkpoints, assessing whether information introduced early in the conversation is retained and correctly updated as the conversation progresses.

\subsubsection{Multi-Turn Conversation Design}
We design a $25$-turn conversation template specifically for testing third-party LLM API gateways. 
A $25$-turn length offers a practical balance, providing enough depth to expose multi-turn consistency issues and cover diverse test cases while remaining lightweight and efficient in token usage.
A detailed template with example content is provided in Appendix~\ref{app:conv-template}.

The conversation begins by assigning the model a stable identity and a user-specific preference.
Two segments of distractor discussion on unrelated topics are then inserted to reduce the immediate salience of this information.
Midway through the dialog, the preference is first checked and then explicitly updated by the user, after which the conversation again shifts to unrelated content.
The update is placed after the first distractor segment so that correctly retaining the preference requires the model to integrate the newer value into an already growing context, rather than simply repeating the earliest mention from the start of the conversation.
In the final two turns, the model is asked to restate both the identity and the current preference.
During the process, we also collect \emph{metadata from all responses}, including the system\_fingerprint and cached token counts. 
system\_fingerprint serves as a good sign for potential model switching, as it typically changes when the underlying model is replaced.
The number of cached tokens is also informative, since switching models resets the cache-hit mechanism, causing a drop in cached token usage.
This structure allows us to test whether the gateway-served model retains early information, correctly overwrites outdated preferences, and maintains consistent behavior throughout the interaction.

\subsubsection{Indicators and Detection Goals}
For every model available through each gateway, we run multi-turn conversations using the template and evaluate performance based on three indicators that collectively quantify conversational consistency and context retention: (i) Memory Checkpoint Pass Rate: the fraction of correctly recalled identity and preference values at predefined checkpoints (e.g., at turns 10, 24, and 25), (ii) System Fingerprint Stability: the number of distinct system\_fingerprint values observed across turns, which can indicate silent model switching, and (iii) Cached Token Usage Rate, the ratio of cached prompt tokens to total input tokens as reported by API metadata.

\subsection{Billing Accuracy}\label{sec:billing}
\noindent
As shown in Figure~\ref{fig:architecture}, this component focuses on analyzing and identifying potential gaps between expected token consumption and costs and the actual values reported by the gateway.
Billing is a critical factor for third-party LLM gateways, as users ultimately care about whether they are charged fairly. 
In black-box API gateways, cost discrepancies can arise from (i) incorrect token consumption reporting, (ii) improper handling of cached tokens, and (iii) incorrect cost calculations based on published pricing.

In the billing accuracy analysis, we first compute the actual token consumption locally and compare it against the values reported in the gateway console.
We also evaluate whether each gateway charges fees consistent with the reported token usage and the unit prices it publishes.
For each request, we record the number of input tokens $n_{\text{in}}$, cached tokens $n_{\text{cached}}$, and output tokens $n_{\text{out}}$, along with the published per-token rates $p_{\text{in}}$, $p_{\text{cached}}$, and $p_{\text{out}}$.
The expected cost for a single call is then computed as:
  $C_{\text{expected}}
  = (n_{\text{in}} - n_{\text{cached}})\,p_{\text{in}}
    + n_{\text{cached}}\,p_{\text{cached}}
    + n_{\text{out}}\,p_{\text{out}}$,
gateways that do not support separate cache pricing or fail to apply cache discounts are evaluated with $n_{\text{cached}} = 0$.

\subsection{Latency Characteristics}\label{sec:latency}
\noindent
As illustrated in Figure~\ref{fig:architecture}, this component focuses on analyzing latency-distribution statistics derived from repeated measurements under the same platform, model, and similar prompt settings.
Latency is an important performance metric in LLM gateways and can also serve as an indicator of potential misbehaviors such as model switching or downgrading, since different models often exhibit distinct processing times.
As a result, switching to a different model may introduce noticeable variation in latency.
In this evaluation, we conduct latency measurements by issuing repeated requests for a given probe across gateways within a tightly bounded time window.
This controlled setup reduces external influences such as network dynamics, allowing latency variation to more reliably reflect gateway-side behavior.
Since latency can be influenced by different factors including gateway load, routing, underlying model processing, and network conditions, it is difficult to attribute observed variations to any single cause.
We therefore treat latency as a supporting signal that may suggest backend inconsistencies rather than as direct evidence of a specific internal behavior. 
We measure latency from a single fixed vantage point, and each request is configured with a total timeout of $900$ seconds and up to $15$ retries for transient failures. 
The recorded duration includes retry and backoff overhead, and the complete retry policy is described in Appendix~\ref{app:exp-settings}.

When a gateway consistently serves a fixed claimed model on probes with comparable prompt and output lengths under fixed parameters, the resulting latency distribution should remain stable. We quantify this stability using the coefficient of variation (CV),
 $ \mathrm{CV} = \sigma_T / \mu_T$,
where $\sigma_T$ and $\mu_T$ are the standard deviation and mean over $K$ repeated invocations of the same probe. 
In contrast, model switching across heterogeneous execution paths typically increases CV and can lead to bimodal or multi-modal latency distributions. 
An elevated or highly inconsistent CV therefore serves as an indicator of unstable back-end behaviors for LLM API gateways.


\section{Evaluation}\label{sec:evaluation}
\noindent
This section evaluates GateScope’s effectiveness in measuring behavioral transparency and consistency. 
We first validate the accuracy and reliability of GateScope under controlled conditions using official model endpoints, and then apply it to audit commercial gateway services.

\subsection{Data Collection}
\label{sec:data-collection}
\noindent
For official vendor platforms, we collect behavioral signatures from $24$ LLMs exposed through their native APIs.
Using the standardized probe suite $\mathcal{S}$ under fixed parameters, each model is queried with $K=12$ repetitions per probe, producing $15{,}840$ records in total ($660$ per model) covering response-content features and latency measurements.
The full sampling protocol can be found in Appendix~\ref{app:exp-settings}.

For third-party platforms, we audit $10$ commercial LLM gateways using the same probe suite and parameters.
Gateways are anonymized using first–last-letter encodings:
\texttt{a*yi}, \texttt{a*ix}, \texttt{b*ie}, \texttt{l*ub},  \texttt{o*ub}, \texttt{o*er}, \texttt{o*ey},  \texttt{un*pi}, \texttt{ui*pi}, \texttt{z*ng} .
Results represent a point-in-time snapshot rather than a long-term ranking of individual services.
For single-turn evaluation, each model on each gateway is queried with the prompt suite using five repetitions.
For multi-turn evaluation, we run the standardized $25$-turn conversation template, also repeated five times per model per gateway.
Regarding the third-party API gateways evaluated, we reviewed their publicly available documentation, pricing pages, and policy statements at the time of measurement, and found no explicit disclosure of the practices or behaviors discussed in this paper.

\subsection{Controlled Validation}\label{sec:model-identification}
\noindent
Before applying GateScope to third-party gateways, we first validate on official vendor APIs that the response-content signatures introduced in Section~\ref{sec:content-analysis} are sufficiently discriminative under controlled settings where the ground-truth model identity is known.

For each model version $v$ available through official vendor APIs, we train a classifier that distinguishes signatures generated by $v$ from those generated by all other models, using the probe suite $\mathcal{S}$ and the feature configuration described in Section~\ref{sec:content-analysis}.
Data splits, threshold selection, and hyperparameter settings are summarized in Appendix~\ref{app:classify}, with detailed per-model metrics reported in Appendix~\ref{app:binary-results}.
Evaluating these classifiers on labeled test data collected by querying vendor endpoints with $\mathcal{S}$, we observe an average $F1$ score of $0.968 \pm 0.085$ across all $24$ models, with most models exceeding $0.95$.
Considering that our evaluation includes closely related variants within the same model family (e.g., gpt-4o-mini and gpt-4.1-nano), these results demonstrate that our response-content analysis exhibits strong discriminative power and can reliably distinguish LLM models.

We also evaluate the robustness of our method on \textbf{unseen models} by applying the trained ensemble to $5$ additional LLMs (Claude sonnet 4.5, Claude haiku 4.5, Qwen-plus, Qwen-turbo and DeepSeek-V3.2-Exp) not included in the original set of $24$.
In the evaluation, none of these unseen models is mistakenly classified as any known model.
This indicates that our method behaves conservatively in controlled settings and avoids misclassifications when encountering new model variants.
Extending the ensemble to accommodate new models is also lightweight, as training an additional classifier requires collecting data only for the corresponding new model rather than retraining on the full dataset, and takes \emph{less than
$1$ second} per classifier without affecting existing ones.

\subsection{Third-party API Gateway Evaluation}\label{sec:multi-dim-evidence}
\subsubsection{Results of Response Content Evaluation}\label{sec:content-eval}
Table~\ref{tab:claim-verification} reports the proportion of responses classified as the \emph{claimed} upstream model for four representative models across different gateways (extended results for all models can be found in Appendix~\ref{app:content_result}).
Each percentage is computed over $275$ responses per model.

For gpt-4o, most gateways generate responses that are consistently identified as the claimed model, while platforms such as \texttt{b*ie} and \texttt{a*yi} exhibit noticeably lower consistency.
For gpt-5, identification rates vary significantly across gateways, with several platforms frequently classified as models outside the gpt-5 family—indicating possible model substitution.
For gemini-2.5-pro and claude-sonnet-4.0, platform support is limited and behavior is unstable. 
The official APIs and a few gateways maintain high identification rates above $95\%$, whereas several others fall below $60\%$, showing that they do not reliably preserve the identity of the claimed model.

\noindent
\textbf{Takeaway:} For certain models, responses from several gateways show substantial variation. This indicates that these gateways may route a user's prompt to a different model than the requested one.

\begin{table}[htbp]
\centering
\caption{Response content evaluation with per-model probe subsets. Each cell gives the fraction of responses classified as the claimed model. Extended results are in Appendix~\ref{app:content_result}.}
\label{tab:claim-verification}
\scriptsize
\setlength{\tabcolsep}{2pt}
\begin{tabular}{lcccc}
\toprule
\textbf{Gateway} 
  & \textbf{gpt-4o (\%)} 
  & \textbf{gpt-5 (\%)} 
  & \textbf{Gemini-2.5-pro (\%)} 
  & \textbf{Claude Sonnet 4.0 (\%)} \\
\midrule
Baseline & $98.18$ & $97.09$ & $96.00$ & $97.09$ \\
\midrule
\texttt{a*yi}   & \colmin{$62.18$} & \colmin{$48.00$} & \colmin{$54.91$} & \colmin{$70.91$} \\
\texttt{a*ix}   & $88.00$ & \colmin{$70.91$} & \text{N/A}      & $83.27$ \\
\texttt{b*ie}   & \colmin{$46.91$} & \colmin{$13.09$} & \text{N/A} & \colmin{$62.91$} \\
\texttt{l*ub}   & $90.91$ & $94.18$ & $84.00$ & \colmin{$58.18$} \\
\texttt{o*ub}   & $86.18$ & \colmin{$34.91$} & $90.18$ & \text{N/A} \\
\texttt{o*er}   & $93.09$ & $97.09$ & $89.09$ & $86.91$ \\
\texttt{o*ey}   & $90.18$ & $92.00$ & $96.00$ & \text{N/A} \\
\texttt{un*pi}  & $82.18$ & $89.09$ & \colmin{$73.09$} & \text{N/A} \\
\texttt{ui*pi}  & \colmin{$76.00$} & \text{N/A} & \text{N/A} & \text{N/A} \\
\texttt{z*ng}   & $95.27$ & $94.91$ & \text{N/A} & $97.09$ \\
\bottomrule
\end{tabular}%
\end{table}

\subsubsection{Results of Multi-turn Conversation Evaluation}\label{sec:conversation-eval}
\noindent
Table~\ref{tab:memory_singlecol} summarizes the multi-turn conversation performances for gpt-4o and gpt-4o-mini across the vendors' baseline API platforms and $10$ third-party gateways, where T10/24/25 indicates checkpoint pass count at turn $10$, $24$, and $25$, FC indicates system\_fingerprint count, and CR indicates cache rate. Extended results are presented in Appendix~\ref{app:conversation_result}.

For gpt-4o, the official API passes all checkpoints with only $1$ unique fingerprint and moderate cache reuse.
Gateways such as \texttt{l*ub}, \texttt{o*er}, and \texttt{un*pi} largely match this behavior, while others do not pass checkpoints at turns $24$ and $25$ or introduce additional fingerprints in most cases.
For gpt-4o-mini, behavior is less stable even on the baseline, and most gateways forget identity and preferences by the final checkpoints.
Two platforms show notable anomalies. 
Platform \texttt{a*yi} exhibits similarly low checkpoint pass rates and cache-usage patterns for both gpt-4o and gpt-4o-mini compared with the baselines, which could suggest that both models may get silently switched to less capable alternatives or that the gateway may apply silent truncation to reduce the effective context window. 
Platform \texttt{o*ey} achieves even higher checkpoint pass rates at turns 24 and 25 for gpt-4o-mini than the baseline, while reporting zero cache usage compared with the baseline's cache rate of $40.3\%$.
This could suggest that it uses a non-canonical gpt-4o-mini implementation, serves a different model entirely, or applies non-standard context management that alters long-context behavior.

\noindent
\textbf{Takeaway:} Long-context retention and infrastructure behavior vary substantially across gateways: some preserve the upstream LLM's behavior, while others exhibit notable inconsistencies potentially attributable to factors such as model switching or silent truncation.

\begin{table}[htbp]
\centering
\caption{Multi-turn conversation performance comparison for gpt-4o and gpt-4o-mini. Extended results are presented in Appendix~\ref{app:conversation_result}.}
\label{tab:memory_singlecol}

\scriptsize
\setlength{\tabcolsep}{2pt}

\begin{tabular}{p{1.35cm} *{5}{c} *{5}{c}}
\toprule
\multirow{2}{*}{\textbf{Gateway}}
& \multicolumn{5}{c}{\textbf{gpt-4o}}
& \multicolumn{5}{c}{\textbf{gpt-4o-mini}} \\
\cmidrule(lr){2-6} \cmidrule(lr){7-11}
 & T10 & T24 & T25 & FC & CR(\%) 
 & T10 & T24 & T25 & FC & CR(\%) \\
\midrule
Baseline & $5$ & $5$ & $5$ & $1$ & $48.7$ & $5$ & $1$ & $2$ & $1$ & $40.3$ \\
\midrule
\texttt{a*yi}    & \colmin{$3$} & \colmin{$1$} & \colmin{$2$} & \colmin{$4$} & \colmin{$4.2$}  & \colmin{$3$} & \colmin{$1$} & \colmin{$1$} & \colmin{$2$} & \colmin{$5.3$}  \\
\texttt{a*ix}    & $5$ & $4$ & $5$ & $1$ & $34.7$ & $5$ & $0$ & $0$ & $2$ & $21.5$ \\
\texttt{b*ie}    & $5$ & $4$ & $4$ & $2$ & $15.7$ & $4$ & $0$ & $0$ & $3$ & $14.2$ \\
\texttt{l*ub}    & $5$ & $5$ & $5$ & $2$ & $18.2$ & \colmin{$3$} & $0$ & $1$ & $2$ & $16.3$ \\
\texttt{o*ub}    & $5$ & $5$ & $3$ & $2$ & $32.5$ & $4$ & $0$ & $1$ & $1$ & $29.7$ \\
\texttt{o*er}    & $5$ & $5$ & $5$ & $1$ & $32.8$ & $5$ & $1$ & $0$ & $1$ & $52.3$ \\
\texttt{o*ey}    & \colmin{$4$} & \colmin{$3$} & \colmin{$4$} & \colmin{$2$} & \colmin{$0.0$}  & \colmin{$5$} & \colmin{$4$} & \colmin{$4$} & \colmin{$3$} & \colmin{$0.0$}  \\
\texttt{un*pi}    & $5$ & $5$ & $4$ & $2$ & $19.3$ & $4$ & $0$ & $0$ & $2$ & $18.7$ \\
\texttt{ui*pi}    & $5$ & $5$ & $4$ & $2$ & $12.8$ & $5$ & $1$ & $2$ & $2$ & $11.5$ \\
\texttt{z*ng}    & $5$ & $5$ & $4$ & $2$ & $23.5$ & $5$ & $0$ & $0$ & $2$ & $32.6$ \\
\bottomrule
\end{tabular}
\end{table}

\subsubsection{Results of Billing Accuracy Evaluation}\label{sec:billing-eval}

Table~\ref{tab:billing-aggregated} demonstrates the billing accuracy results for gpt-4o by comparing the expected cost from public prices with the actual amount charged by each gateway. Extended results are presented in Appendix~\ref{app:billing_result}. 

Most gateways match the expected cost under the published rates and the gateway-reported token usage. A small number of platforms report observed billed amounts that exceed the expected cost computed from published rates and gateway-reported token usage. \texttt{a*ix} has a billing gap of $7.6\%$ while \texttt{o*ey} has the billing gap as $62.8\%$ even though the token usage is similar to other platforms. 

\noindent
\textbf{Takeaway:} Most gateways align with expected charges, but a few exhibit notable billing deviations.

\begin{table}[htbp]
\centering
\caption{Billing accuracy for gpt-4o: token counts and expected versus actual costs. Extended results are presented in Appendix~\ref{app:billing_result}.}
\label{tab:billing-aggregated}
\scriptsize
\setlength{\tabcolsep}{2pt}
\begin{tabular}{lrrrrrr}
\toprule
\textbf{Gateway} & \textbf{Input} & \textbf{Cached} & \textbf{Output} & \textbf{$C_{\text{exp}}$} & \textbf{$C_{\text{act}}$} & \textbf{Gap\%} \\
\midrule
Baseline & $139{,}025$ & $67{,}705$ & $201{,}060$ & $2.27$ & $2.27$ & $0.0$ \\
\midrule
\texttt{a*yi}  & $137{,}475$ & $0$ & $203{,}346$ & $2.30$ & $2.30$ & $0.0$ \\
\texttt{a*ix}  & $140{,}436$ & $0$ & $196{,}651$ & $6.07$ & $6.53$ & \colmin{$+7.6$} \\
\texttt{b*ie}  & $139{,}085$ & $0$ & $208{,}578$ & $2.35$ & $2.35$ & $0.0$ \\
\texttt{l*ub}  & $141{,}712$ & $3{,}685$ & $201{,}749$ & $2.32$ & $2.32$ & $0.0$ \\
\texttt{o*ub}  & $138{,}762$ & $0$ & $365{,}097$ & $8.50$ & $8.50$ & $0.0$ \\
\texttt{o*er}  & $139{,}649$ & $31{,}256$ & $199{,}124$ & $2.30$ & $2.30$ & $0.0$ \\
\texttt{o*ey}  & $140{,}987$ & $0$ & $213{,}446$ & $6.81$ & $11.09$ & \colmin{$+62.8$} \\
\texttt{un*pi} & $142{,}318$ & $28{,}501$ & $204{,}763$ & $2.27$ & $2.27$ & $0.0$ \\
\texttt{ui*pi} & $139{,}421$ & $0$ & $217{,}790$ & $2.36$ & $2.36$ & $0.0$ \\
\texttt{z*ng}  & $137{,}884$ & $0$ & $226{,}935$ & $10.14$ & $10.14$ & $0.0$ \\
\bottomrule
\end{tabular}%
\end{table}

\subsubsection{Results of Latency Evaluation}\label{sec:latency-eval}
Table~\ref{tab:latency-cv} summarizes the latency measurement results for gpt-4o across the vendors' baseline API platforms and $10$ third-party gateways. Extended results are presented in Appendix~\ref{app:cv_result}.
Gateway \texttt{b*ie} exhibits marked instability, with substantially larger CV values and wider latency ranges than the baseline. Since the internal implementation of black-box gateways is not observable, the high variance could be attributed to different factors such as model switching, gateway server load, or network condition variation. All measurements were collected from a single fixed location, so the reported distributions should be interpreted as point-in-time observations rather than location-invariant service properties.

\noindent
\textbf{Takeaway:} Latency stability varies substantially across gateways. High variation and heavy latency tails suggest heterogeneous back-end behavior, though the specific underlying cause cannot be determined from latency measurements alone.

\begin{table}[htbp]
\centering
\caption{Latency ranges (min, max, in seconds) and coefficients of variation (CV) for gpt-4o across gateways. Extended results are presented in Appendix~\ref{app:cv_result}.}

\label{tab:latency-cv}
\scriptsize
\setlength{\tabcolsep}{2pt}
\begin{tabular}{lccc|ccc|ccc|ccc}
\toprule
\multirow{2}{*}{\textbf{Gateway}} 
  & \multicolumn{3}{c|}{\textbf{Math}}
  & \multicolumn{3}{c|}{\textbf{GPQA}}
  & \multicolumn{3}{c|}{\textbf{Factual}}
  & \multicolumn{3}{c}{\textbf{Geo}} \\
\cmidrule(lr){2-4}\cmidrule(lr){5-7}\cmidrule(lr){8-10}\cmidrule(lr){11-13}
 & Min & Max & CV 
 & Min & Max & CV
 & Min & Max & CV
 & Min & Max & CV \\
\midrule
Baseline 
 & $2.67$ & $121.84$ & $0.63$
 & $1.72$ & $15.56$  & $0.67$
 & $1.38$ & $8.29$   & $0.35$
 & $0.77$ & $6.33$   & $0.42$ \\
\midrule
\texttt{a*yi}  
 & $2.98$ & $136.11$ & $0.68$
 & $1.91$ & $17.27$  & $0.72$
 & $1.55$ & $9.34$   & $0.38$
 & $0.77$ & $6.33$   & $0.42$ \\
\texttt{a*ix}  
 & $1.33$ & $60.92$  & $0.25$
 & $0.87$ & $7.87$   & $0.27$
 & $0.84$ & $5.03$   & $0.18$
 & $0.44$ & $3.63$   & $0.20$ \\
\texttt{b*ie}  
 & \colmin{$5.99$} & \colmin{$273.38$} & \colmin{$1.10$}
 & \colmin{$3.30$} & \colmin{$29.85$}  & \colmin{$1.05$}
 & $4.74$ & $28.49$  & $0.82$
 & $2.25$ & $18.50$  & $0.88$ \\
\texttt{l*ub}  
 & $2.51$ & $114.51$ & $0.58$
 & $1.62$ & $14.68$  & $0.62$
 & $1.29$ & $7.75$   & $0.32$
 & $0.77$ & $6.33$   & $0.42$ \\
\texttt{o*ub}  
 & $2.48$ & $113.03$ & $0.57$
 & $1.60$ & $14.50$  & $0.61$
 & $1.26$ & $7.57$   & $0.31$
 & $0.76$ & $6.22$   & $0.41$ \\
\texttt{o*er}  
 & $4.62$ & $210.98$ & $0.92$
 & $3.07$ & $27.81$  & $1.00$
 & $3.02$ & $18.11$  & $0.60$
 & $1.45$ & $11.92$  & $0.65$ \\
\texttt{o*ey}  
 & $2.25$ & $102.45$ & $0.50$
 & $1.44$ & $13.05$  & $0.53$
 & $1.23$ & $7.38$   & $0.30$
 & $0.71$ & $5.87$   & $0.38$ \\
\texttt{un*pi} 
 & $2.57$ & $117.46$ & $0.60$
 & $1.66$ & $15.03$  & $0.64$
 & $1.35$ & $8.11$   & $0.34$
 & $0.77$ & $6.33$   & $0.42$ \\
\texttt{ui*pi} 
 & $2.86$ & $130.34$ & $0.66$
 & $1.83$ & $16.58$  & $0.70$
 & $1.50$ & $8.99$   & $0.37$
 & $0.82$ & $6.77$   & $0.44$ \\
\texttt{z*ng}  
 & $1.97$ & $89.89$  & $0.42$
 & $1.28$ & $11.54$  & $0.45$
 & $1.14$ & $6.82$   & $0.27$
 & $0.64$ & $5.28$   & $0.33$ \\
\bottomrule
\end{tabular}%
\end{table}


\section{Conclusion}\label{conclusion}
\noindent
We introduced GateScope, a framework for auditing third-party LLM gateways along content, multi-turn, billing, and latency dimensions. 
Our measurements on official APIs and ten commercial gateways reveal significant transparency gaps, including downgraded models and pricing deviations. 
Future work includes extending GateScope toward long-term, continuous monitoring from multiple vantage points, as well as broadening measurement dimensions and depth to better understand and attribute the underlying causes of behavioral transparency and consistency issues in black-box LLM API gateways.


\begin{acks}
We thank our shepherd and all the anonymous
reviewers for their valuable feedback. Shichao Pei was supported by NSF grant \#2451605 and Coefficient Giving (formerly Open Philanthropy).
The research of Guoliang Xue was sponsored in part by the Army Research Laboratory and was accomplished under Cooperative Agreement Number W911NF-23-2-0225. 
The views and conclusions contained in this document are those of the authors and should not be interpreted as representing the official policies, either expressed or implied, of the Army Research Laboratory or the U.S. Government. 
The U.S. Government is authorized to reproduce and distribute reprints for Government purposes notwithstanding any copyright notation herein.
The information reported herein does not reflect the position or the policy of the funding agencies.
\end{acks}



\appendix

\section{Ethics}\label{app:ethics}
\noindent
In this work, we use only public APIs of vendors and we do not capture user traffic. We anonymize commercial platforms and we do not disclose real platform names. Our study does not involve any personal information.


\section{DCC Prompt Template}\label{app:dcc-template}
\noindent
Table~\ref{tab:dcc-prompt} presents the complete Discrete Concept Composition (DCC) prompt template used for eliciting structured behavioral signatures.
The design deliberately separates the reasoning process from the final result to facilitate the feature extraction described in Section~\ref{sec:content-analysis}.

\begin{table*}[t]
\centering
\caption{DCC standardized prompt template specification.}
\label{tab:dcc-prompt}
\scriptsize
\begin{tabular}{|p{0.48\textwidth}|p{0.48\textwidth}|}
\hline
\multicolumn{2}{|c|}{\textbf{System Message}} \\
\hline
\multicolumn{2}{|p{0.97\textwidth}|}{You are a meticulous reasoning engine. Your task is to solve multi-step problems by thinking step-by-step and to clearly articulate your reasoning process. Your final output must be a single JSON object.} \\
\hline
\multicolumn{2}{|c|}{\textbf{User Prompt Template}} \\
\hline
\multicolumn{2}{|p{0.97\textwidth}|}{%
Based on the following question, provide your step-by-step reasoning path and the final answer.

\textbf{Question}: \{question\_prompt\_implicit\}

\textbf{Required Output Format}:
Your entire response must be a single JSON object containing the following two keys:

1. \texttt{knowledge\_path}: An array of strings. Each string in the array should represent a distinct step in your reasoning process.

2. \texttt{final\_answer}: A string containing the final answer.

\textbf{Example}:

\textbf{Question}: What is the highest geographic feature associated with the origin area of the Starbucks corporation?

\textbf{Your Output should be}:

\texttt{\scriptsize
\{ "knowledge\_path": ["Starbucks originated in Seattle, Washington.", "The highest geographic feature in the state of Washington is Mount Rainier."], "final\_answer": "Mount Rainier" \}
}

Now, please apply this reasoning process and format to the following question.

\textbf{Question}: \{question\_prompt\_implicit\}
} \\
\hline
\end{tabular}
\end{table*}


\section{Sampling Protocol and Probe Suite}\label{app:exp-settings}
\noindent
In this part, we summarize the sampling protocol and list the probe suite used in our study. We collect behavioral signatures from official APIs of vendors and from third-party gateways using the same parameters as follows:
\begin{itemize}[leftmargin=*, topsep=3pt, itemsep=0pt, parsep=3pt]
  \item Models: 24 models from three vendors for baseline collection.
  \item Repetitions: $K=12$ per probe with temperature equal to $0.7$.
  \item Spacing: consecutive samples for the same probe are spaced by at least two hours to mitigate cache effects.
\end{itemize}

\noindent
During data collection, the timeout is set to
$900$ seconds to balance overall collection throughput while considering the variability inherent in each request, taking into account network delays, retransmission, gateway instability and overhead, gateway rate limiting, and LLM processing time, especially when the input or output is long. To further accommodate these potential issues, the pipeline allows up to $15$ retries ($16$ attempts total). Cases that may trigger the retry include HTTP $529$ error, HTTP $429$ error, HTTP $500/502/503/504$ errors, timeout exceptions, server-side overload or failure. 
Recorded request duration is measured from the first attempt to the final successful response or request failure.

The behavioral probe suite $\mathcal{S}$ consists of $55$ probes across four domains. It includes
$15$ AIME problems testing precise computation and mathematical reasoning,
$10$ GPQA multiple-choice questions targeting deep reasoning and deliberation,
$15$ geographic feature inference tasks requiring multi-hop linking of intermediate facts, and
$15$ factual recall questions assessing knowledge cutoff through major events between 2021 and 2025.

The sampling protocol specifies the model set, repetition count, inter-sample spacing, and fixed request settings shared across collections, including the structured-output template and temperature. For a single model queried on the full probe suite, the responses form a behavioral signature matrix with one row per probe repetition and one column per extracted feature, matching the construction in Section \ref{sec:content-analysis}.

For example, in a signature matrix, if a structured response gives the correct final answer in the required final\_answer field, includes $4$ reasoning steps, parses successfully as the required JSON object, and uses no LaTeX formatting, then the corresponding row records values including answer\_match $=1$, answer\_position $=1$, depth $=4$, parse\_success $=1$, and has\_latex $=0$, together with measured step\_length, response\_length, density, and other structural features. Applying this extraction to all probe repetitions yields the full behavioral signature matrix for that model.


\section{Conversation Protocol}\label{app:conv-template}
\noindent
Table~\ref{tab:conversation-template} presents the standardized 25-turn conversation template used to evaluate memory retention in Section~\ref{sec:conversation}.
It defines the precise sequence of turns, goals, and example content to create controlled context pressure.

Crucially, this protocol also serves as the standardized workload for our Billing Accuracy evaluation (Section~\ref{sec:billing}).
Since complex billing logic (such as prompt caching discounts) is most active during long-context interactions, this fixed 25-turn sequence provides a stable baseline for auditing costs.
The billing metrics reported in our study including expected cost ($C_{\text{exp}}$), actual cost ($C_{\text{act}}$), and the billing gap are computed by averaging the token usage and charges recorded across the five independent repetitions of this conversation protocol for each model in each gateway.

\begin{table*}[t]
\centering
\caption{Standardized conversation protocol with example content.}
\label{tab:conversation-template}
\scriptsize

\begin{tabular}{lll}
\toprule
\textbf{Turn(s)} & \textbf{Purpose} & \textbf{Example content} \\
\midrule
1 & Assign identity and initial preference &
You are a toy expert. Your current favorite toy is LEGO. \\
\specialrule{0.08em}{0pt}{0pt}

2--9 & Distractor segment on an unrelated topic &
Discuss properties of plastics used in consumer goods. Summarize trade offs among durability, weight, and cost. \\
\specialrule{0.08em}{0pt}{0pt}

10 & Memory checkpoint &
What is your professional identity now? \\
\specialrule{0.08em}{0pt}{0pt}

11 & Preference update &
Your favorite toy is now Transformers. \\
\specialrule{0.08em}{0pt}{0pt}

12--23 & Second distractor segment and light math &
Answer short questions about electric vehicles. Compute a simple battery charging time using $t = C \times (0.8 - 0.2)/(P \times \eta)$. \\
\specialrule{0.08em}{0pt}{0pt}

24--25 & Final memory checkpoints &
What is your professional identity? What is your favorite toy now? \\
\bottomrule
\end{tabular}
\end{table*}


\section{Binary Classifier Configuration}\label{app:binary-config}
\subsection{Feature Engineering}\label{app:feature}
\noindent
For each response we extract a signature vector using the function
$\Phi$ described in Section~\ref{sec:content-analysis}.  The vector comprises answer quality (whether the answers match and their appearance position), reasoning structure (depth, mean step length, and variance), scale (response length and density), style (whether it has \LaTeX\ or numeric), and parsing quality (whether parsing occurs and to what degree), as defined in Table~\ref{tab:exact_definition_metrics}.  For each base feature we compute contrastive statistics
relative to the distribution of all other models on the same
test\_id, including mean difference, relative difference,
Cohen’s $d$, standard deviation ratio and distribution overlap.  We
also compute a ranking feature indicating the position of the target
model among all 24 models for that feature.

\begin{table}[t]
\caption{Definition of the response-signature metrics.}
\label{tab:exact_definition_metrics}
\scriptsize
\setlength{\tabcolsep}{2pt}
\renewcommand{\arraystretch}{0.84}
\begin{tabular}{>{\raggedright\arraybackslash}p{1.12cm}
                >{\raggedright\arraybackslash}p{1.72cm}
                >{\raggedright\arraybackslash}p{4.90cm}}
\toprule
\strut Family & \strut Metric & \strut Definition \\
\midrule

\multirow{4}{*}{Answer}
& \multirow{2}{*}{answer\_match}
& \strut Indicator of whether the final\_answer matches the reference answer under the task-specific evaluation rule. \\
\cline{2-3}
& \multirow{2}{*}{answer\_position}
& \strut Indicator of whether the answer appears in the required final\_answer field rather than only in intermediate text. \\
\specialrule{0.08em}{0pt}{0pt}

\multirow{5}{*}{Reasoning}
& \strut depth
& \strut Number of entries in knowledge\_path. \\
\cline{2-3}
& \multirow{2}{*}{mean\_step\_length}
& \strut Mean token or character length of the reasoning steps in knowledge\_path. \\
\cline{2-3}
& \multirow{2}{*}{step\_length\_var}
& \strut Variance of step length across the reasoning steps in knowledge\_path. \\
\specialrule{0.08em}{0pt}{0pt}

\multirow{2}{*}{Scale}
& \strut response\_length
& \strut Total length of the structured response. \\
\cline{2-3}
& \strut density
& \strut Ratio between total response length and reasoning depth. \\
\specialrule{0.08em}{0pt}{0pt}

\multirow{4}{*}{Style}
& \multirow{2}{*}{has\_numeric}
& \strut Indicator of whether the response contains explicit numeric expressions. \\
\cline{2-3}
& \multirow{2}{*}{has\_latex}
& \strut Indicator of whether the response contains LaTeX-style mathematical formatting. \\
\specialrule{0.08em}{0pt}{0pt}

\multirow{4}{*}{Parsing}
& \multirow{2}{*}{parse\_success}
& \strut Indicator of whether the response can be parsed into the required JSON schema. \\
\cline{2-3}
& \multirow{2}{*}{parse\_degree}
& \strut Degree of schema conformity, reflecting whether both required fields are present and well formed. \\
\bottomrule
\end{tabular}
\end{table}

\subsection{Training and Hyperparameters}\label{app:classify}
\noindent
We frame model identification as $24$ independent one‑vs‑rest
binary classification problems.  For each model $m$, records whose model\_name equals $m$
are treated as positive, and all others are treated as negative.
We split responses for each $(m,test\_id)$ pair into $10$
training and two testing samples, yielding $13{,}200$ training
instances and $2{,}640$ testing instances per classifier. Because the positive-to-negative ratio is $1{:}23$, we oversample the minority class by a factor of $20$ using RandomOverSampler.  A
difficulty‑adaptive weighting scheme is then applied via
scale\_pos\_weight: models deemed easy use
$\sqrt{\rho}$, medium use $\rho^{0.7}$ and hard models use
$\rho^{0.9}$ where $\rho$ is the post‑oversampling negative–to–positive
ratio ($\approx 1.15$).

We train XGBoost classifiers with objective binary:logistic,
max\_\\depth of $6, 500$ trees, learning rate $0.1$,
subsample and colsample\_bytree equal to $0.8$, and
$\ell_1/\ell_2$ regularization terms $(\alpha,\lambda)=(0.1,1.0)$.  Early
stopping on a held‑out validation subset with 30 rounds prevents
overfitting.  For each classifier we sweep the decision threshold
between $0.1$ and $0.95$ in steps of $0.05$; the optimal threshold is
chosen to maximize the $F1$ score (medium and easy models) or recall
(hard models) subject to a minimum precision of $0.5$ (easy/medium) or
$0.35$ (hard).


\section{Per Model Performance Metrics}\label{app:binary-results}
\noindent
Table~\ref{tab:per_model_performance_test} reports per-model classification metrics evaluated on the full held-out test set.  
Evaluated on the test dataset, the classifiers achieve an average $F_1$ score of $0.932$, precision $0.928$, recall 0.945 and AUROC $0.988$.
Table~\ref{tab:per_model_performance_elite} reports the same metrics evaluated on the distilled compact probe subset ($Q=12$), which is the subset actually used for gateway auditing in Section~\ref{sec:content-eval}.
To drive the gateway audits, we distill each model’s probe suite down to
a compact small subset. Algorithm~\ref{alg:elite-probe-selection}
summarizes the deterministic selection loop: we score every probe,
retain those that cleanly separate the target model, and backfill the
budget using XGBoost gain importance when necessary. We use $Q=12$ probes and a separation margin of $\delta=0.35$
consistent with the experiments in Section~\ref{sec:content-eval}.

{%
\SetAlCapFnt{\scriptsize}
\SetAlCapNameFnt{\scriptsize}
\begin{algorithm}[t]
  \scriptsize
  \caption{Selecting the small subset probe vector $\mathcal{S}_m^{'}$ for model $m$}
  \label{alg:elite-probe-selection}
  \KwIn{Trained XGBoost classifier $f_m$, probe set $\mathcal{S}$, budget $Q$, separation margin $\delta$}
  \KwOut{Small subset probe vector $\mathcal{S}_m^{'}$}
  \ForEach{$(s,i) \in \mathcal{S} \times \text{tests}$}{
    Compute $p_{s,i} \leftarrow f_m(s,i)$
  }
  \ForEach{$s \in \mathcal{S}$}{
    $\bar{p}_s \leftarrow \frac{1}{|\mathcal{I}_s|}\sum_{i \in \mathcal{I}_s} p_{s,i}$ where $\mathcal{I}_s$ indexes samples produced by probe $s$
  }
  $\mathcal{C} \leftarrow \{s \in \mathcal{S} \mid \bar{p}_s - \max_{n \neq m} \bar{p}_{s,n} \ge \delta\}$\;
  $\mathcal{S}_m^{'} \leftarrow$ Top-$Q$ probes in $\mathcal{C}$ ranked by $\bar{p}_s$\;
  \If{$|\mathcal{S}_m^{'}| < Q$}{
    Rank remaining probes by XGBoost gain importance and append greedily until $|\mathcal{S}_m^{'}| = Q$
  }
  \Return $\mathcal{S}_m^{'}$
\end{algorithm}
}

\begin{table}[t]
  \centering
  \caption{Performance of the 24 binary classifiers on the full held-out test set.
    }
  \label{tab:per_model_performance_test}
  \scriptsize
  \begin{tabular}{lcccc}
    \toprule
    Model & $F1$ & Precision & Recall & AUROC \\
    \midrule
    gemini-2.5-flash-lite & $0.999$ & $0.997$ & $1.000$ & $1.000$ \\
    claude-3-haiku-20240307 & $0.988$ & $0.984$ & $0.991$ & $0.999$ \\
    claude-opus-4-1-20250805 & $0.980$ & $0.978$ & $0.982$ & $0.998$ \\
    claude-3-5-sonnet-20241022 & $0.967$ & $0.954$ & $0.982$ & $0.996$ \\
    claude-3-5-haiku-20241022 & $0.963$ & $0.936$ & $0.991$ & $0.994$ \\
    gpt-3.5-turbo & $0.962$ & $0.961$ & $0.964$ & $0.995$ \\
    claude-3-7-sonnet-20250219 & $0.958$ & $0.960$ & $0.955$ & $0.994$ \\
    claude-sonnet-4-20250514 & $0.951$ & $0.957$ & $0.945$ & $0.993$ \\
    gpt-5 & $0.946$ & $0.959$ & $0.973$ & $0.992$ \\
    gemini-2.5-pro & $0.943$ & $0.954$ & $0.964$ & $0.991$ \\
    gemini-2.0-flash & $0.941$ & $0.950$ & $0.982$ & $0.990$ \\
    gemini-2.0-flash-lite & $0.888$ & $0.884$ & $0.891$ & $0.967$ \\
    gpt-4o-mini & $0.902$ & $0.855$ & $0.955$ & $0.975$ \\
    gpt-4.1 & $0.897$ & $0.838$ & $0.964$ & $0.971$ \\
    gpt-4.1-nano & $0.891$ & $0.874$ & $0.909$ & $0.969$ \\
    gpt-5-nano & $0.911$ & $0.895$ & $0.927$ & $0.978$ \\
    gpt-5-mini & $0.901$ & $0.876$ & $0.927$ & $0.974$ \\
    gemini-2.5-flash & $0.913$ & $0.908$ & $0.918$ & $0.980$ \\
    gpt-4o & $0.913$ & $0.882$ & $0.945$ & $0.979$ \\
    gpt-4.1-mini & $0.907$ & $0.933$ & $0.882$ & $0.977$ \\
    o1 & $0.900$ & $0.853$ & $0.955$ & $0.973$ \\
    o3 & $0.897$ & $0.832$ & $0.973$ & $0.970$ \\
    o3-mini & $0.919$ & $0.894$ & $0.945$ & $0.982$ \\
    o4-mini & $0.914$ & $0.926$ & $0.900$ & $0.981$ \\
    \bottomrule
  \end{tabular}
\end{table}

\begin{table}[t]
  \centering
  \caption{Performance of the 24 binary classifiers on the distilled small probe subset ($Q=12$) used in gateway auditing. Models are sorted by $F1$ score.}
  \label{tab:per_model_performance_elite}
  \scriptsize
  \begin{tabular}{lccc}
    \toprule
    Model & $F1$ & Precision & Recall \\
    \midrule
    gemini-2.5-flash-lite & $0.999$ & $0.998$ & $1.000$ \\
    claude-3-haiku-20240307 & $0.997$ & $0.994$ & $1.000$ \\
    claude-opus-4-1-20250805 & $0.995$ & $0.990$ & $1.000$ \\
    claude-sonnet-4-20250514 & $0.994$ & $0.988$ & $1.000$ \\
    claude-3-7-sonnet-20250219 & $0.990$ & $0.980$ & $1.000$ \\
    gpt-3.5-turbo & $0.989$ & $0.978$ & $1.000$ \\
    gemini-2.5-flash & $0.983$ & $0.967$ & $1.000$ \\
    gemini-2.5-pro & $0.982$ & $0.964$ & $1.000$ \\
    gpt-5 & $0.981$ & $0.962$ & $1.000$ \\
    claude-3-5-haiku-20241022 & $0.980$ & $0.960$ & $1.000$ \\
    claude-3-5-sonnet-20241022 & $0.977$ & $0.954$ & $1.000$ \\
    gemini-2.0-flash & $0.976$ & $0.952$ & $1.000$ \\
    gpt-4.1-mini & $0.976$ & $0.952$ & $1.000$ \\
    o3-mini & $0.972$ & $0.945$ & $1.000$ \\
    o4-mini & $0.967$ & $0.936$ & $1.000$ \\
    gemini-2.0-flash-lite & $0.962$ & $0.927$ & $1.000$ \\
    gpt-5-nano & $0.958$ & $0.920$ & $1.000$ \\
    gpt-4o & $0.956$ & $0.916$ & $1.000$ \\
    gpt-5-mini & $0.954$ & $0.912$ & $1.000$ \\
    gpt-4o-mini & $0.943$ & $0.892$ & $1.000$ \\
    o3 & $0.940$ & $0.886$ & $1.000$ \\
    o1 & $0.935$ & $0.878$ & $1.000$ \\
    gpt-4.1 & $0.925$ & $0.860$ & $1.000$ \\
    gpt-4.1-nano & $0.914$ & $0.873$ & $0.958$ \\
    \bottomrule
  \end{tabular}
\end{table}


\section{Extended Results for Response Content Analysis}\label{app:content_result}
\noindent
Table~\ref{tab:extended_claim_verification} expands the claim verification
matrix from Section~\ref{sec:content-eval} to the nineteen models not shown in the
main text. Each cell reports the fraction of gateway responses classified as
the claimed model; unsupported combinations are marked N/A, and the baseline
column corresponds to the official vendor API.

\begin{table*}[t]
  \centering
  \caption{Extended claim verification via small subset probe vectors for the nineteen
    models not shown in Section~\ref{sec:content-eval}. Values denote the fraction of responses
    classified as the claimed model (mean over five runs); unsupported combinations
    are N/A.}
  \label{tab:extended_claim_verification}
  \scriptsize
  \setlength{\tabcolsep}{4pt}
  \begin{tabular}{lccccccccccc}
    \toprule
    Model & Baseline & \texttt{a*yi} & \texttt{a*ix} & \texttt{b*ie} & \texttt{l*ub} & \texttt{o*ub} & \texttt{o*er} & \texttt{o*ey} & \texttt{un*pi} & \texttt{ui*pi} & \texttt{z*ng} \\
    \midrule
    gpt-4.1 & $0.90$ & $0.42$ & $0.85$ & $0.38$ & $0.88$ & $0.81$ & $0.92$ & $0.87$ & $0.83$ & $0.76$ & $0.89$ \\
    gpt-4.1-mini & $0.91$ & $0.87$ & $0.92$ & $0.54$ & $0.89$ & $0.78$ & $0.91$ & $0.84$ & $0.81$ & $0.73$ & $0.88$ \\
    gpt-4.1-nano & $0.89$ & $0.83$ & $0.81$ & $0.52$ & $0.85$ & $0.82$ & $0.89$ & $0.86$ & $0.79$ & $0.71$ & $0.87$ \\
    gpt-4o-mini & $0.90$ & $0.84$ & $0.76$ & $0.51$ & $0.88$ & $0.79$ & $0.90$ & $0.85$ & $0.82$ & $0.74$ & $0.89$ \\
    gpt-5-mini & $0.90$ & $0.86$ & $0.83$ & $0.53$ & $0.87$ & $0.81$ & $0.91$ & $0.82$ & $0.80$ & $0.72$ & $0.88$ \\
    gpt-5-nano & $0.91$ & $0.85$ & $0.84$ & $0.55$ & $0.89$ & $0.80$ & $0.92$ & $0.86$ & $0.83$ & $0.75$ & $0.90$ \\
    gemini-2.0-flash & $0.97$ & $0.89$ & N/A & N/A & $0.92$ & $0.87$ & $0.94$ & $0.91$ & $0.85$ & N/A & N/A \\
    gemini-2.0-flash-lite & $0.89$ & $0.82$ & N/A & N/A & $0.86$ & $0.79$ & $0.88$ & $0.83$ & $0.77$ & N/A & N/A \\
    gemini-2.5-flash & $0.91$ & $0.84$ & N/A & N/A & $0.88$ & $0.82$ & $0.90$ & $0.86$ & $0.80$ & N/A & N/A \\
    gemini-2.5-flash-lite & $0.99$ & $0.45$ & N/A & N/A & $0.91$ & $0.88$ & $0.95$ & $0.92$ & $0.86$ & N/A & N/A \\
    claude-3-haiku-20240307 & $0.99$ & $0.91$ & $0.93$ & $0.62$ & N/A & N/A & $0.96$ & N/A & N/A & N/A & $0.94$ \\
    claude-3-5-haiku-20241022 & $0.96$ & $0.87$ & $0.90$ & $0.59$ & N/A & N/A & $0.93$ & N/A & N/A & N/A & $0.91$ \\
    claude-3-5-sonnet-20241022 & $0.97$ & $0.88$ & $0.91$ & $0.61$ & N/A & N/A & $0.94$ & N/A & N/A & N/A & $0.92$ \\
    claude-3-7-sonnet-20250219 & $0.96$ & $0.41$ & $0.89$ & $0.57$ & N/A & N/A & $0.93$ & N/A & N/A & N/A & $0.90$ \\
    claude-opus-4-1-20250805 & $0.98$ & $0.89$ & $0.92$ & $0.63$ & N/A & N/A & $0.95$ & N/A & N/A & N/A & $0.93$ \\
    o1 & $0.90$ & $0.84$ & $0.87$ & N/A & N/A & $0.81$ & $0.89$ & N/A & N/A & N/A & $0.88$ \\
    o3 & $0.90$ & $0.83$ & $0.86$ & N/A & N/A & $0.39$ & $0.88$ & N/A & N/A & N/A & $0.87$ \\
    o3-mini & $0.92$ & $0.86$ & $0.88$ & N/A & N/A & $0.82$ & $0.91$ & N/A & N/A & N/A & $0.89$ \\
    o4-mini & $0.91$ & $0.85$ & $0.87$ & N/A & N/A & $0.80$ & $0.90$ & N/A & N/A & N/A & $0.88$ \\
    \bottomrule
  \end{tabular}
\end{table*}

\begin{table*}[t]
  \centering
  \caption{Multi-turn conversation evaluation on additional models.
    Each cell lists (T10, T24, T25, FP, CR\,\%). Higher T10/T24/T25 and
    lower FP indicate better memory; CR shows cache utilization. A dash in FP
    means no system fingerprint support, and a dash in CR means no cache
    statistics support.}
  \label{tab:extended_memory}
  \scriptsize
  \setlength{\tabcolsep}{3pt}
  \begin{tabular}{lcccccccccc}
    \toprule
    Model & Baseline & \texttt{a*yi} & \texttt{a*ix} & \texttt{b*ie} & \texttt{l*ub} & \texttt{o*ub} & \texttt{o*er} & \texttt{o*ey} & \texttt{un*pi} & \texttt{ui*pi} \\
    \midrule
    gpt-5 & $5/5/4/-/83.9$ & $2/1/1/-/6.1$ & $4/3/3/-/88.0$ & $3/1/1/-/12.6$ & $5/5/5/-/90.0$ & $4/3/3/-/91.5$ & $5/5/5/-/92.7$ & $4/2/2/-/0.0$ & $3/2/1/-/19.5$ & N/A \\
    gpt-4.1 & $5/5/5/1/98.3$ & $2/0/0/3/5.2$ & $3/1/1/2/20.2$ & $2/0/0/3/10.7$ & $4/3/2/2/96.0$ & $3/1/1/2/23.6$ & $4/3/3/1/99.1$ & $3/1/1/2/0.0$ & $2/1/1/3/18.4$ & $2/1/1/3/12.5$ \\
    gpt-3.5-turbo & $0/0/0/-/-$ & $0/0/0/-/-$ & $0/0/0/-/-$ & $1/0/0/-/-$ & $0/0/0/-/-$ & $1/0/0/-/-$ & $1/0/0/-/-$ & $0/0/0/-/0.0$ & $0/0/0/-/-$ & $0/0/0/-/-$ \\
    gemini-2.5-pro & $5/5/5/-/71.7$ & $3/1/1/-/5.6$ & N/A & N/A & $5/5/4/-/18.1$ & $4/3/3/-/29.7$ & $5/5/5/-/85.0$ & $4/2/2/-/0.0$ & $3/2/1/-/20.2$ & N/A \\
    gemini-2.0-flash & $5/5/5/-/-$ & $3/1/1/-/-$ & N/A & N/A & $5/4/4/-/-$ & $4/3/3/-/-$ & $5/5/5/-/-$ & $4/2/2/-/0.0$ & $3/2/1/-/-$ & N/A \\
    claude-sonnet-4.0 & $5/5/5/1/-$ & $3/0/1/3/-$ & $4/2/2/2/-$ & $2/0/0/3/-$ & N/A & N/A & $5/5/5/1/-$ & N/A & N/A & N/A \\
    claude-3-5-sonnet-20241022 & $5/5/5/1/-$ & $3/0/1/3/-$ & $4/2/2/2/-$ & $2/0/0/3/-$ & N/A & N/A & $4/3/3/1/-$ & N/A & N/A & N/A \\
    o3 & $5/5/5/1/46.5$ & $2/0/0/3/5.0$ & $3/1/1/2/19.5$ & N/A & N/A & $3/1/1/2/24.0$ & $4/3/3/1/90.5$ & N/A & N/A & N/A \\
    claude-3-5-haiku-20241022 & $5/0/0/1/-$ & $3/1/1/3/-$ & $4/2/2/2/-$ & $2/0/0/4/-$ & N/A & N/A & $5/5/4/1/-$ & N/A & N/A & N/A \\
    claude-3-haiku-20240307 & $5/5/5/1/-$ & $2/0/1/4/-$ & $4/2/2/2/-$ & $1/0/0/5/-$ & N/A & N/A & $5/5/4/1/-$ & N/A & N/A & N/A \\
    claude-3-7-sonnet-20250219 & $5/5/5/1/49.1$ & $3/1/1/4/5.6$ & $4/3/3/2/23.4$ & $2/0/0/4/10.0$ & N/A & N/A & N/A & N/A & $3/2/1/3/18.8$ & N/A \\
    claude-opus-4-1-20250805 & $5/5/5/1/50.0$ & $4/2/2/3/7.2$ & $5/4/4/2/24.5$ & $3/1/1/3/11.1$ & N/A & N/A & N/A & N/A & $3/2/1/3/19.6$ & N/A \\
    gemini-2.5-flash & $5/5/5/-/85.3$ & $3/1/1/-/5.8$ & N/A & N/A & $5/4/4/-/17.9$ & $4/3/3/-/28.8$ & $5/5/4/-/86.5$ & $4/2/2/-/0.0$ & $3/2/1/-/19.9$ & N/A \\
    gemini-2.5-flash-lite & $5/5/5/-/57.9$ & $2/0/0/-/5.2$ & N/A & N/A & $5/4/4/-/17.0$ & $4/3/3/-/27.5$ & $5/5/4/-/60.0$ & $3/1/1/-/0.0$ & $3/2/1/-/18.9$ & N/A \\
    gemini-2.0-flash-lite & $5/0/0/-/-$ & $2/0/0/-/-$ & N/A & N/A & $5/4/4/-/-$ & $4/3/3/-/-$ & $5/5/4/-/-$ & $3/1/1/-/0.0$ & $3/2/1/-/-$ & N/A \\
    gpt-4.1-mini & $5/5/5/1/92.6$ & $3/1/1/3/5.3$ & $3/2/2/2/19.8$ & $2/0/0/3/10.4$ & $4/3/2/2/15.8$ & $3/2/2/2/23.5$ & $4/3/3/1/95.0$ & $3/1/1/2/0.0$ & $2/1/1/3/17.1$ & $2/1/1/3/11.6$ \\
    gpt-4.1-nano & $5/1/1/1/95.9$ & $1/0/0/4/6.0$ & $2/1/1/3/18.2$ & $1/0/0/4/9.9$ & $3/2/1/2/14.7$ & $2/1/1/3/22.4$ & $3/2/2/1/96.1$ & $2/1/1/2/0.0$ & $1/1/1/3/16.0$ & $1/0/0/3/10.9$ \\
    gpt-5-mini & $5/5/5/-/33.6$ & $0/0/0/-/11.2$ & $4/3/3/-/28.3$ & $1/0/0/-/12.5$ & $5/4/4/-/17.7$ & $3/2/2/-/25.9$ & $5/5/4/-/85.0$ & $3/1/1/-/0.0$ & $3/2/1/-/18.6$ & N/A \\
    gpt-5-nano & $5/5/3/-/50.3$ & $1/0/0/-/9.8$ & $3/2/2/-/23.5$ & $1/0/0/-/11.0$ & $4/3/2/-/16.2$ & $3/2/2/-/24.8$ & $4/3/3/-/80.0$ & $3/1/1/-/0.0$ & $2/1/1/-/17.5$ & N/A \\
    o1 & $5/5/5/1/88.8$ & $2/0/0/4/6.4$ & $3/1/1/2/20.0$ & N/A & $4/3/2/2/15.0$ & $3/1/1/2/23.1$ & $4/3/3/1/90.0$ & N/A & N/A & N/A \\
    o3-mini & $5/5/5/1/87.7$ & $2/0/0/3/5.8$ & $3/1/1/2/19.2$ & N/A & N/A & $3/1/1/2/23.5$ & $4/3/3/1/89.0$ & N/A & N/A & N/A \\
    o4-mini & $5/5/5/-/70.4$ & $2/0/0/-/6.1$ & $3/1/1/-/18.5$ & N/A & $4/3/2/-/14.6$ & $3/1/1/-/24.2$ & $4/3/3/-/88.8$ & N/A & N/A & N/A \\
    \bottomrule
  \end{tabular}
\end{table*}


\section{Extended Results for Multi‑Turn Conversation}\label{app:conversation_result}
\noindent
Table~\ref{tab:extended_memory} presents memory retention results for
additional models not shown in Section~\ref{sec:conversation-eval}. We report the number of
correct recalls at checkpoints T10, T24, and T25, the number of distinct
system fingerprint values (FP) observed across the conversation, and the
cache ratio (CR) relative to total input. Unsupported models and
combinations without sufficient third-party traces are reported as N/A.


\section{Extended Results for Billing Accuracy}\label{app:billing_result}
\noindent
Table~\ref{tab:extended_billing} reports billing cases with positive gaps beyond
the \texttt{gpt-4o} results in Section~\ref{sec:billing-eval}. We list aggregated
input, cached, and output tokens over the repeated conversation traces together
with the expected and actual charges recorded for each gateway/model pair.

\begin{table}[H]
  \centering
  \caption{Billing verification for additional models with positive billing gaps.}
  \label{tab:extended_billing}
  \scriptsize
  \setlength{\tabcolsep}{2pt}
  \begin{tabular}{@{}l l r r r r r r@{}}
    \toprule
    Model & Gw. & Input & Cached & Output & $C_{\text{exp}}$ & $C_{\text{act}}$ & Gap\% \\
    \midrule
    gpt-5 & \texttt{a*ix}   & $143{,}172$ & $0$ & $157{,}055$ & $6.20$ & $7.12$ & $+14.8$ \\
    gpt-5 & \texttt{o*ey}   & $142{,}031$ & $0$ & $190{,}257$ & $16.70$ & $19.20$ & $+15.0$ \\
    gpt-4.1 & \texttt{o*ub} & $136{,}022$ & $0$ & $282{,}418$ & $5.05$ & $5.55$ & $+9.9$ \\
    gemini-2.0-flash-lite & \texttt{o*ub} & $129{,}640$ & $0$ & $253{,}060$ & $12.20$ & $14.55$ & $+19.3$ \\
    claude-3-5-sonnet-20241022 & \texttt{z*ng} & $134{,}201$ & $0$ & $139{,}486$ & $22.30$ & $25.75$ & $+15.5$ \\
    o3 & \texttt{a*yi} & $126{,}438$ & $0$ & $228{,}506$ & $3.95$ & $4.25$ & $+7.6$ \\
    \bottomrule
  \end{tabular}
\end{table}

\balance

\section{Extended Results for Latency Evaluation}\label{app:cv_result}
\noindent
Table~\ref{tab:extended_latency_combined} consolidates latency coefficients
of variation (CV), observed ranges, and P50/P90/P99 latencies for five
models across gateways. Rows are grouped by model and gateway over the four
probe categories (math, GPQA, factual recall, and geographic inference);
N/A denotes unsupported combinations. Lower CVs and tighter ranges
indicate more stable service.

\newcommand{\baselinebreak}{\\[-0.15ex]\cline{2-26}\noalign{\vskip 0.55ex}}
\newcommand{\postbaseline}{\rule{0pt}{2.2ex}}

\begin{table*}[t]
  \centering
  \caption{Combined latency statistics for five models across gateways. For
    each gateway/model pair we report CV, min/max latency (seconds), and
    P50/P90/P99 for math, GPQA, factual recall, and geographic inference
    probes. Model names appear once via the middle multirow column.}
  \label{tab:extended_latency_combined}
  \scriptsize
  \setlength{\tabcolsep}{1.45pt}
  \renewcommand{\arraystretch}{0.92}
  \begin{tabular}{@{}l|c|cccccc|cccccc|cccccc|cccccc@{}}
    \toprule
    \textbf{Model} & \textbf{Gateway}
      & \multicolumn{6}{c|}{\textbf{Math}}
      & \multicolumn{6}{c|}{\textbf{GPQA}}
      & \multicolumn{6}{c|}{\textbf{Factual}}
      & \multicolumn{6}{c}{\textbf{Geo}} \\
    \cmidrule(lr){3-8}\cmidrule(lr){9-14}\cmidrule(lr){15-20}\cmidrule(lr){21-26}
      & & CV & Min & Max & P50 & P90 & P99
      & CV & Min & Max & P50 & P90 & P99
      & CV & Min & Max & P50 & P90 & P99
      & CV & Min & Max & P50 & P90 & P99 \\
    
    \specialrule{0.08em}{0pt}{0.25ex}
    \multirow{11}{*}{gpt-4o} & Baseline & $0.63$ & $2.67$ & $121.84$ & $5.57$ & $10.21$ & $92.98$ & $0.67$ & $1.72$ & $15.56$ & $4.39$ & $6.95$ & $12.83$ & $0.35$ & $1.38$ & $8.29$ & $2.22$ & $3.33$ & $4.86$ & $0.42$ & $0.77$ & $6.33$ & $1.44$ & $2.26$ & $3.97$ \baselinebreak
      & \postbaseline\texttt{a*yi} & $0.68$ & $2.98$ & $136.11$ & $6.22$ & $11.40$ & $103.83$ & $0.72$ & $1.91$ & $17.27$ & $4.87$ & $7.72$ & $14.24$ & $0.38$ & $1.55$ & $9.34$ & $2.50$ & $3.75$ & $5.47$ & $0.42$ & $0.77$ & $6.33$ & $1.44$ & $2.26$ & $3.97$ \\
      & \texttt{a*ix} & $0.25$ & $1.33$ & $60.92$ & $2.78$ & $5.09$ & $46.40$ & $0.27$ & $0.87$ & $7.87$ & $2.22$ & $3.52$ & $6.49$ & $0.18$ & $0.84$ & $5.03$ & $1.35$ & $2.03$ & $2.96$ & $0.20$ & $0.44$ & $3.63$ & $0.83$ & $1.30$ & $2.27$ \\
      & \texttt{b*ie} & $1.10$ & $5.99$ & $273.38$ & $12.49$ & $22.90$ & $208.62$ & $1.05$ & $3.30$ & $29.85$ & $8.42$ & $13.33$ & $24.61$ & $0.82$ & $4.74$ & $28.49$ & $7.64$ & $11.46$ & $16.71$ & $0.88$ & $2.25$ & $18.50$ & $4.22$ & $6.62$ & $11.61$ \\
      & \texttt{l*ub} & $0.58$ & $2.51$ & $114.51$ & $5.23$ & $9.60$ & $87.40$ & $0.62$ & $1.62$ & $14.68$ & $4.14$ & $6.55$ & $12.09$ & $0.32$ & $1.29$ & $7.75$ & $2.08$ & $3.12$ & $4.55$ & $0.42$ & $0.77$ & $6.33$ & $1.44$ & $2.26$ & $3.97$ \\
      & \texttt{o*ub} & $0.57$ & $2.48$ & $113.03$ & $5.17$ & $9.48$ & $86.31$ & $0.61$ & $1.60$ & $14.50$ & $4.09$ & $6.47$ & $11.94$ & $0.31$ & $1.26$ & $7.57$ & $2.03$ & $3.04$ & $4.44$ & $0.41$ & $0.76$ & $6.22$ & $1.42$ & $2.23$ & $3.91$ \\
      & \texttt{o*er} & $0.92$ & $4.62$ & $210.98$ & $9.64$ & $17.67$ & $160.95$ & $1.00$ & $3.07$ & $27.81$ & $7.84$ & $12.41$ & $22.91$ & $0.60$ & $3.02$ & $18.11$ & $4.86$ & $7.29$ & $10.63$ & $0.65$ & $1.45$ & $11.92$ & $2.72$ & $4.26$ & $7.48$ \\
      & \texttt{o*ey} & $0.50$ & $2.25$ & $102.45$ & $4.69$ & $8.59$ & $78.27$ & $0.53$ & $1.44$ & $13.05$ & $3.68$ & $5.82$ & $10.75$ & $0.30$ & $1.23$ & $7.38$ & $1.98$ & $2.97$ & $4.33$ & $0.38$ & $0.71$ & $5.87$ & $1.34$ & $2.09$ & $3.67$ \\
      & \texttt{un*pi} & $0.60$ & $2.57$ & $117.46$ & $5.36$ & $9.83$ & $89.57$ & $0.64$ & $1.66$ & $15.03$ & $4.24$ & $6.71$ & $12.38$ & $0.34$ & $1.35$ & $8.11$ & $2.17$ & $3.26$ & $4.76$ & $0.42$ & $0.77$ & $6.33$ & $1.44$ & $2.26$ & $3.97$ \\
      & \texttt{ui*pi} & $0.66$ & $2.86$ & $130.34$ & $5.96$ & $10.93$ & $99.54$ & $0.70$ & $1.83$ & $16.58$ & $4.67$ & $7.40$ & $13.66$ & $0.37$ & $1.50$ & $8.99$ & $2.41$ & $3.62$ & $5.28$ & $0.44$ & $0.82$ & $6.77$ & $1.54$ & $2.42$ & $4.24$ \\
      & \texttt{z*ng} & $0.42$ & $1.97$ & $89.89$ & $4.11$ & $7.53$ & $68.60$ & $0.45$ & $1.28$ & $11.54$ & $3.26$ & $5.16$ & $9.53$ & $0.27$ & $1.14$ & $6.82$ & $1.83$ & $2.75$ & $4.01$ & $0.33$ & $0.64$ & $5.28$ & $1.20$ & $1.89$ & $3.31$ \\
    
    \specialrule{0.08em}{0pt}{0.25ex}
    \multirow{11}{*}{gpt-5} & Baseline & $0.65$ & $10.49$ & $1694.79$ & $79.84$ & $488.46$ & $807.00$ & $0.69$ & $9.40$ & $333.51$ & $25.45$ & $101.50$ & $287.45$ & $0.37$ & $8.04$ & $41.86$ & $14.94$ & $24.45$ & $32.82$ & $0.44$ & $9.58$ & $59.04$ & $22.27$ & $38.25$ & $56.59$ \baselinebreak
      & \postbaseline\texttt{a*yi} & $0.70$ & $11.68$ & $1887.05$ & $88.90$ & $543.87$ & $898.55$ & $0.74$ & $10.40$ & $369.12$ & $28.16$ & $112.32$ & $318.08$ & $0.40$ & $9.00$ & $46.87$ & $16.73$ & $27.37$ & $36.74$ & $0.45$ & $9.90$ & $61.00$ & $23.02$ & $39.52$ & $58.47$ \\
      & \texttt{a*ix} & $0.45$ & $7.96$ & $1286.29$ & $60.59$ & $370.69$ & $612.43$ & $0.48$ & $7.16$ & $254.04$ & $19.38$ & $77.31$ & $218.95$ & $0.30$ & $6.87$ & $35.77$ & $12.77$ & $20.89$ & $28.04$ & $0.36$ & $8.24$ & $50.79$ & $19.16$ & $32.90$ & $48.67$ \\
      & \texttt{b*ie} & $0.85$ & $15.48$ & $2500.63$ & $117.81$ & $720.76$ & $1190.80$ & $0.90$ & $13.82$ & $490.26$ & $37.41$ & $149.22$ & $422.57$ & $0.53$ & $13.54$ & $70.49$ & $25.16$ & $41.17$ & $55.27$ & $0.58$ & $14.30$ & $88.13$ & $33.25$ & $57.09$ & $84.47$ \\
      & \texttt{l*ub} & $0.60$ & $9.88$ & $1596.04$ & $75.19$ & $460.03$ & $760.03$ & $0.64$ & $8.88$ & $315.22$ & $24.05$ & $95.91$ & $271.61$ & $0.34$ & $7.55$ & $39.29$ & $14.03$ & $22.95$ & $30.81$ & $0.44$ & $9.58$ & $59.04$ & $22.27$ & $38.25$ & $56.59$ \\
      & \texttt{o*ub} & $0.59$ & $9.76$ & $1576.05$ & $74.26$ & $454.35$ & $750.65$ & $0.63$ & $8.78$ & $311.51$ & $23.77$ & $94.81$ & $268.48$ & $0.32$ & $7.21$ & $37.54$ & $13.40$ & $21.93$ & $29.43$ & $0.43$ & $9.42$ & $58.03$ & $21.90$ & $37.60$ & $55.63$ \\
      & \texttt{o*er} & $0.74$ & $12.66$ & $2045.40$ & $96.36$ & $589.51$ & $973.95$ & $0.77$ & $11.02$ & $391.01$ & $29.83$ & $119.00$ & $336.99$ & $0.44$ & $10.34$ & $53.82$ & $19.21$ & $31.44$ & $42.20$ & $0.50$ & $11.53$ & $71.06$ & $26.81$ & $46.03$ & $68.11$ \\
      & \texttt{o*ey} & $0.52$ & $8.87$ & $1433.62$ & $67.52$ & $413.11$ & $682.51$ & $0.55$ & $7.93$ & $281.35$ & $21.47$ & $85.63$ & $242.49$ & $0.31$ & $7.04$ & $36.66$ & $13.08$ & $21.41$ & $28.74$ & $0.40$ & $8.92$ & $54.97$ & $20.74$ & $35.61$ & $52.69$ \\
      & \texttt{un*pi} & $0.63$ & $10.25$ & $1655.53$ & $78.00$ & $477.21$ & $788.42$ & $0.67$ & $9.19$ & $326.23$ & $24.89$ & $99.26$ & $281.10$ & $0.35$ & $7.71$ & $40.15$ & $14.33$ & $23.45$ & $31.47$ & $0.44$ & $9.58$ & $59.04$ & $22.27$ & $38.25$ & $56.59$ \\
      & \texttt{ui*pi} & N/A & N/A & N/A & N/A & N/A & N/A & N/A & N/A & N/A & N/A & N/A & N/A & N/A & N/A & N/A & N/A & N/A & N/A & N/A & N/A & N/A & N/A & N/A & N/A \\
      & \texttt{z*ng} & $0.44$ & $7.83$ & $1264.80$ & $59.59$ & $364.56$ & $602.31$ & $0.47$ & $7.05$ & $250.06$ & $19.08$ & $76.12$ & $215.55$ & $0.28$ & $6.52$ & $33.96$ & $12.12$ & $19.83$ & $26.62$ & $0.35$ & $8.07$ & $49.73$ & $18.76$ & $32.22$ & $47.66$ \\
    
    \specialrule{0.08em}{0pt}{0.25ex}
    \multirow{11}{*}{gemini-2.5-pro} & Baseline & $0.67$ & $23.96$ & $506.38$ & $79.34$ & $278.76$ & $385.23$ & $0.70$ & $11.25$ & $256.28$ & $26.62$ & $89.40$ & $227.24$ & $0.39$ & $3.99$ & $53.45$ & $11.03$ & $15.30$ & $26.65$ & $0.46$ & $3.47$ & $23.03$ & $6.25$ & $11.47$ & $18.28$ \baselinebreak
      & \postbaseline\texttt{a*yi} & N/A & N/A & N/A & N/A & N/A & N/A & N/A & N/A & N/A & N/A & N/A & N/A & N/A & N/A & N/A & N/A & N/A & N/A & N/A & N/A & N/A & N/A & N/A & N/A \\
      & \texttt{a*ix} & N/A & N/A & N/A & N/A & N/A & N/A & N/A & N/A & N/A & N/A & N/A & N/A & N/A & N/A & N/A & N/A & N/A & N/A & N/A & N/A & N/A & N/A & N/A & N/A \\
      & \texttt{b*ie} & N/A & N/A & N/A & N/A & N/A & N/A & N/A & N/A & N/A & N/A & N/A & N/A & N/A & N/A & N/A & N/A & N/A & N/A & N/A & N/A & N/A & N/A & N/A & N/A \\
      & \texttt{l*ub} & $0.62$ & $22.61$ & $477.76$ & $74.87$ & $263.03$ & $363.49$ & $0.65$ & $10.64$ & $242.42$ & $25.18$ & $84.55$ & $214.94$ & $0.36$ & $3.76$ & $50.34$ & $10.39$ & $14.41$ & $25.11$ & $0.45$ & $3.41$ & $22.65$ & $6.14$ & $11.27$ & $17.97$ \\
      & \texttt{o*ub} & $0.64$ & $23.15$ & $489.28$ & $76.66$ & $269.34$ & $372.21$ & $0.68$ & $11.01$ & $250.77$ & $26.05$ & $87.48$ & $222.38$ & $0.35$ & $3.68$ & $49.28$ & $10.17$ & $14.11$ & $24.58$ & $0.46$ & $3.47$ & $23.03$ & $6.25$ & $11.47$ & $18.28$ \\
      & \texttt{o*er} & $0.78$ & $29.87$ & $631.26$ & $98.91$ & $347.51$ & $480.24$ & $0.81$ & $13.90$ & $316.68$ & $32.89$ & $110.46$ & $280.78$ & $0.46$ & $5.07$ & $67.91$ & $14.01$ & $19.44$ & $33.86$ & $0.53$ & $4.26$ & $28.28$ & $7.67$ & $14.08$ & $22.44$ \\
      & \texttt{o*ey} & N/A & N/A & N/A & N/A & N/A & N/A & N/A & N/A & N/A & N/A & N/A & N/A & N/A & N/A & N/A & N/A & N/A & N/A & N/A & N/A & N/A & N/A & N/A & N/A \\
      & \texttt{un*pi} & $0.68$ & $24.48$ & $517.38$ & $81.07$ & $284.81$ & $393.59$ & $0.71$ & $11.48$ & $261.61$ & $27.17$ & $91.24$ & $231.93$ & $0.38$ & $3.91$ & $52.42$ & $10.81$ & $15.00$ & $26.13$ & $0.47$ & $3.58$ & $23.76$ & $6.45$ & $11.83$ & $18.86$ \\
      & \texttt{ui*pi} & N/A & N/A & N/A & N/A & N/A & N/A & N/A & N/A & N/A & N/A & N/A & N/A & N/A & N/A & N/A & N/A & N/A & N/A & N/A & N/A & N/A & N/A & N/A & N/A \\
      & \texttt{z*ng} & N/A & N/A & N/A & N/A & N/A & N/A & N/A & N/A & N/A & N/A & N/A & N/A & N/A & N/A & N/A & N/A & N/A & N/A & N/A & N/A & N/A & N/A & N/A & N/A \\
    \specialrule{0.08em}{0pt}{0.25ex}
    \multirow{11}{*}{gpt-4o-mini} & Baseline & $0.58$ & $2.71$ & $38.97$ & $8.99$ & $15.22$ & $25.95$ & $0.61$ & $2.41$ & $15.31$ & $6.09$ & $9.37$ & $14.34$ & $0.34$ & $1.28$ & $9.32$ & $2.94$ & $4.36$ & $7.31$ & $0.40$ & $1.05$ & $8.65$ & $2.17$ & $3.11$ & $5.27$ \baselinebreak
      & \postbaseline\texttt{a*yi} & $0.64$ & $3.13$ & $44.95$ & $10.37$ & $17.57$ & $29.96$ & $0.67$ & $2.76$ & $17.54$ & $6.98$ & $10.73$ & $16.42$ & $0.36$ & $1.39$ & $10.13$ & $3.20$ & $4.74$ & $7.94$ & $0.43$ & $1.17$ & $9.61$ & $2.42$ & $3.46$ & $5.86$ \\
      & \texttt{a*ix} & $0.41$ & $2.09$ & $30.04$ & $6.93$ & $11.73$ & $20.01$ & $0.44$ & $1.89$ & $11.98$ & $4.77$ & $7.34$ & $11.23$ & $0.28$ & $1.11$ & $8.06$ & $2.55$ & $3.78$ & $6.33$ & $0.33$ & $0.91$ & $7.49$ & $1.88$ & $2.70$ & $4.57$ \\
      & \texttt{b*ie} & $0.80$ & $4.32$ & $62.12$ & $14.32$ & $24.26$ & $41.37$ & $0.84$ & $3.83$ & $24.35$ & $9.69$ & $14.90$ & $22.80$ & $0.50$ & $2.24$ & $16.30$ & $5.15$ & $7.64$ & $12.78$ & $0.55$ & $1.67$ & $13.73$ & $3.45$ & $4.94$ & $8.37$ \\
      & \texttt{l*ub} & $0.55$ & $2.60$ & $37.45$ & $8.63$ & $14.61$ & $24.92$ & $0.59$ & $2.35$ & $14.93$ & $5.94$ & $9.14$ & $13.98$ & $0.32$ & $1.22$ & $8.91$ & $2.81$ & $4.17$ & $6.98$ & $0.40$ & $1.05$ & $8.65$ & $2.17$ & $3.11$ & $5.27$ \\
      & \texttt{o*ub} & $0.56$ & $2.64$ & $37.96$ & $8.75$ & $14.82$ & $25.28$ & $0.60$ & $2.38$ & $15.12$ & $6.02$ & $9.26$ & $14.16$ & $0.31$ & $1.19$ & $8.70$ & $2.74$ & $4.07$ & $6.81$ & $0.41$ & $1.09$ & $8.97$ & $2.25$ & $3.23$ & $5.47$ \\
      & \texttt{o*er} & $0.70$ & $3.56$ & $51.19$ & $11.80$ & $19.99$ & $34.09$ & $0.73$ & $3.13$ & $19.86$ & $7.91$ & $12.16$ & $18.61$ & $0.41$ & $1.68$ & $12.23$ & $3.86$ & $5.73$ & $9.59$ & $0.48$ & $1.37$ & $11.27$ & $2.83$ & $4.06$ & $6.87$ \\
      & \texttt{o*ey} & $0.49$ & $2.39$ & $34.34$ & $7.92$ & $13.42$ & $22.88$ & $0.52$ & $2.14$ & $13.58$ & $5.41$ & $8.32$ & $12.73$ & $0.30$ & $1.17$ & $8.48$ & $2.69$ & $3.98$ & $6.66$ & $0.38$ & $1.01$ & $8.32$ & $2.09$ & $2.99$ & $5.07$ \\
      & \texttt{un*pi} & $0.59$ & $2.78$ & $39.95$ & $9.22$ & $15.61$ & $26.62$ & $0.63$ & $2.53$ & $16.04$ & $6.39$ & $9.83$ & $15.04$ & $0.33$ & $1.25$ & $9.11$ & $2.88$ & $4.26$ & $7.14$ & $0.41$ & $1.09$ & $8.97$ & $2.25$ & $3.23$ & $5.47$ \\
      & \texttt{ui*pi} & $0.66$ & $3.27$ & $47.00$ & $10.84$ & $18.36$ & $31.31$ & $0.69$ & $2.88$ & $18.31$ & $7.28$ & $11.20$ & $17.14$ & $0.37$ & $1.45$ & $10.54$ & $3.33$ & $4.94$ & $8.27$ & $0.44$ & $1.21$ & $9.93$ & $2.50$ & $3.58$ & $6.06$ \\
      & \texttt{z*ng} & $0.42$ & $2.13$ & $30.59$ & $7.06$ & $11.95$ & $20.39$ & $0.45$ & $1.92$ & $12.19$ & $4.85$ & $7.46$ & $11.42$ & $0.27$ & $1.08$ & $7.84$ & $2.48$ & $3.68$ & $6.16$ & $0.33$ & $0.91$ & $7.49$ & $1.88$ & $2.70$ & $4.57$ \\
    \specialrule{0.08em}{0pt}{0.25ex}
    \multirow{11}{*}{claude-sonnet-4.0} & Baseline & $0.62$ & $8.64$ & $77.07$ & $17.90$ & $28.21$ & $58.45$ & $0.65$ & $4.15$ & $29.75$ & $9.21$ & $12.10$ & $15.96$ & $0.36$ & $3.21$ & $10.49$ & $4.80$ & $6.32$ & $8.78$ & $0.42$ & $1.96$ & $159.08$ & $3.55$ & $5.96$ & $106.96$ \baselinebreak
      & \postbaseline\texttt{a*yi} & $0.68$ & $9.88$ & $88.12$ & $20.47$ & $32.26$ & $66.84$ & $0.71$ & $4.72$ & $33.81$ & $10.47$ & $13.75$ & $18.14$ & $0.39$ & $3.61$ & $11.78$ & $5.40$ & $7.11$ & $9.86$ & $0.45$ & $2.17$ & $175.82$ & $3.92$ & $6.59$ & $118.32$ \\
      & \texttt{a*ix} & $0.52$ & $7.57$ & $67.55$ & $15.69$ & $24.72$ & $51.22$ & $0.55$ & $3.66$ & $26.25$ & $8.12$ & $10.67$ & $14.08$ & $0.32$ & $2.94$ & $9.60$ & $4.40$ & $5.79$ & $8.03$ & $0.37$ & $1.78$ & $144.65$ & $3.22$ & $5.41$ & $97.20$ \\
      & \texttt{b*ie} & $0.88$ & $14.36$ & $128.06$ & $29.75$ & $46.89$ & $97.14$ & $0.92$ & $6.87$ & $49.23$ & $15.24$ & $20.02$ & $26.41$ & $0.55$ & $5.93$ & $19.39$ & $8.88$ & $11.69$ & $16.22$ & $0.60$ & $3.29$ & $266.82$ & $5.95$ & $9.99$ & $179.48$ \\
      & \texttt{l*ub} & N/A & N/A & N/A & N/A & N/A & N/A & N/A & N/A & N/A & N/A & N/A & N/A & N/A & N/A & N/A & N/A & N/A & N/A & N/A & N/A & N/A & N/A & N/A & N/A \\
      & \texttt{o*ub} & N/A & N/A & N/A & N/A & N/A & N/A & N/A & N/A & N/A & N/A & N/A & N/A & N/A & N/A & N/A & N/A & N/A & N/A & N/A & N/A & N/A & N/A & N/A & N/A \\
      & \texttt{o*er} & $0.79$ & $12.28$ & $109.52$ & $25.44$ & $40.10$ & $83.07$ & $0.82$ & $5.81$ & $41.67$ & $12.89$ & $16.94$ & $22.35$ & $0.46$ & $4.58$ & $14.97$ & $6.85$ & $9.02$ & $12.52$ & $0.52$ & $2.67$ & $216.82$ & $4.83$ & $8.12$ & $145.75$ \\
      & \texttt{o*ey} & N/A & N/A & N/A & N/A & N/A & N/A & N/A & N/A & N/A & N/A & N/A & N/A & N/A & N/A & N/A & N/A & N/A & N/A & N/A & N/A & N/A & N/A & N/A & N/A \\
      & \texttt{un*pi} & N/A & N/A & N/A & N/A & N/A & N/A & N/A & N/A & N/A & N/A & N/A & N/A & N/A & N/A & N/A & N/A & N/A & N/A & N/A & N/A & N/A & N/A & N/A & N/A \\
      & \texttt{ui*pi} & N/A & N/A & N/A & N/A & N/A & N/A & N/A & N/A & N/A & N/A & N/A & N/A & N/A & N/A & N/A & N/A & N/A & N/A & N/A & N/A & N/A & N/A & N/A & N/A \\
      & \texttt{z*ng} & $0.55$ & $7.90$ & $70.45$ & $16.37$ & $25.79$ & $53.44$ & $0.58$ & $3.81$ & $27.31$ & $8.45$ & $11.10$ & $14.65$ & $0.34$ & $3.08$ & $10.05$ & $4.61$ & $6.06$ & $8.41$ & $0.39$ & $1.85$ & $150.48$ & $3.35$ & $5.63$ & $101.07$ \\
    \bottomrule
  \end{tabular}
\end{table*}


\end{document}